\documentclass[prd,floatfix,nofootinbib,preprintnumbers]{revtex4}
\usepackage{epsfig}
\usepackage{graphicx}
\usepackage{amsfonts}
\usepackage{mathrsfs}

\begin{document}

\newcommand{\be}{\begin{equation}}
\newcommand{\ee}{\end{equation}}
\newcommand{\bea}{\begin{eqnarray}}
\newcommand{\eea}{\end{eqnarray}}
\newcommand{\nnb}{\nonumber}
\renewcommand{\thefootnote}{\fnsymbol{footnote}}
\def\lsim{\raise0.3ex\hbox{$\;<$\kern-0.75em\raise-1.1ex\hbox{$\sim\;$}}}
\def\gsim{\raise0.3ex\hbox{$\;>$\kern-0.75em\raise-1.1ex\hbox{$\sim\;$}}}
\def\Frac#1#2{\frac{\displaystyle{#1}}{\displaystyle{#2}}}
\def\no{\nonumber\\}
\def\slash#1{\ooalign{\hfil/\hfil\crcr$#1$}}
\def\ep{\eta^{\prime}}
\def\susy{\mbox{\tiny SUSY}}
\def\sm{\mbox{\tiny SM}}
\def\pslash{\rlap{\hspace{0.02cm}/}{p}}
\def\qslash{\rlap{/}{q}}
\def\kslash{\rlap{\hspace{0.02cm}/}{k}}
\def\lslash{\rlap{\hspace{0.011cm}/}{\ell}}
\def\nslash{\rlap{\hspace{0.02cm}/}{n}}
\def\Pslash{\rlap{\hspace{0.065cm}/}{P}}
\textheight      250mm  

\vskip0.5pc

\title{CP asymmetries in $B\to \phi
K_S$ and $B\to \eta' K_S$ in MSSM}
\author{Jian-Feng Cheng$^{a,b}$,  Chao-Shang Huang$^a$, and Xiao-Hong Wu$^{c,d}$}
\affiliation{
 $^a$ Institute of Theoretical Physics, Academia Sinica, P. O. Box 2735,
             Beijing 100080,  China\\
 $^b$ Institute of High Energy Physics, Academia Sinica, P. O. Box 918(4),
             Beijing 100039,  China\\
 $^c$ Department of Physics, KAIST, Daejeon 305-701, Korea \\
 $^d$ Department of Physics, Peking University, Beijing 100871, China }

\begin{abstract}
We study the $B\to \phi K_S$ and $B\to \ep K_S$ decays in MSSM by
calculating hadronic matrix elements of operators with QCD
factorization approach and including neutral Higgs boson (NHB)
contributions. We calculate the Wilson coefficients of operators
including the new operators which are induced by NHB penguins at
LO using the MIA with double insertions. We calculate the
$\alpha_s$ order hadronic matrix elements of the new operators for
$B\rightarrow \phi K_s$ and $B\rightarrow \eta^\prime K_s$. It is
shown that the recent experimental results on the time-dependent
CP asymmetries in $B\to \phi K_S$ and $B\to \ep K_S$, $S_{\phi K}$
is negative and $S_{\eta^{\prime} K}$ is positive, which can not
be explained in SM, can be explained in MSSM if there are flavor
non-diagonal squark mass matrix elements of 2nd and 3rd
generations whose size satisfies all relevant constraints from
known experiments ($B\to X_S\gamma, B_s\to \mu^+\mu^-, B\to X_s
\mu^+\mu^-, B\to X_s g, \Delta M_s$, etc.). In particular, we find
that one can explain the experimental results with a flavor
non-diagonal mass insertion of chirality LL or LR or RR when
$\alpha_s$ corrections of hadronic matrix elements of operators
are included, in contrast with the claim in the literature. At the
same time, the branching ratios for the two decays can also be in
agreement with experimental measurements.
\end{abstract}

\maketitle
\noindent

\section{Introduction}
The measurements of the time dependent CP asymmetry $S_{J/\psi K}$
in $B\to J/\psi K_S$ have established the presence of CP
violation in neutral B meson decays and the measured
value\cite{sj} \be S_{J/\psi K}=\sin (2 \beta (J/\psi
K_{S}))_{\rm world-ave}=0.734 \pm 0.054. \label{sjp}\ee is in
agreement with the prediction in the standard model (SM).
Recently, various measurements of  CP violation in B factory
experiments have attracted much interest. Among
them\cite{2002,2003}, \begin{eqnarray} S_{\phi K_S}&=&-0.39\pm
0.41,\,\,\,\,\,2002\,\,\, {\rm World-average}\nnb\\S_{\phi
K_S}&=&-0.15\pm 0.33,\,\,\,\,\, 2003\,\,\, {\rm World-average}
\end{eqnarray} is
especially interesting since it deviates greatly from the SM
expectation \be S_{\phi K_S}=\sin (2 \beta (\phi K_S))= \sin (2
\beta (J/\psi K_S)\!) \! + \! O(\lambda^2 \!)
\end{equation}
where $\lambda \simeq 0.2$ appears in Wolfenstein's
parameterization of the CKM matrix. Obviously, the impact of these
experimental results on the validity of CKM and SM is currently
statistically limited. However, they have attracted much interest in
searching for new physics~\cite{dat,kk,kane,chw,hz} and it has been
shown that the deviation can be understood without contradicting
the smallness of the SUSY effect on $B \to J/\psi K_S$ in the
minimal supersymmetric standard model (MSSM)~\cite{kane,chw}.

Another experimental result, which is worth to notice, is of the
time dependent CP asymmetry $S_{\eta^{\prime} K_S}$ in $B\to
\eta^\prime K_S$\cite{eta,2003} \begin{eqnarray}
S_{\eta^\prime K_S}&=& 0.02 \pm 0.34 \pm 0.03 \hspace{5mm}{\rm BaBar}\nonumber\\
&=& 0.43 \pm 0.27 \pm 0.05 \hspace{5mm}{\rm Belle}
\end{eqnarray} which deviates sizably from the SM
expectation. Although both the asymmetries $S_{\phi K}$ and
$S_{\eta^\prime K}$ are smaller than the SM value, $S_{\phi K}$ is
negative and $S_{\eta^\prime K}$ is positive. Because the quark
subprocess $b\to s\bar{s}s$ contributes to both $B\to \phi K_S$
and $B\to \eta^\prime K_S$ decays one should simultaneously
explain the experimental data in a model with the same parameters.
It has been done in Ref.\cite{kkou} in a model- independent way in
the supersymmetric (SUSY) framework. In Ref.\cite{kkou} the
analysis is carried out using the naive factorization to calculate
hadronic matrix elements of operators and the neutral Higgs boson
(NHB) contributions are not included. As we have shown in a
letter\cite{chw} that both the branching ratio (Br) and CP
asymmetry are significantly dependent of the $\alpha_s$
corrections of hadronic matrix elements and NHB contributions are
important in MSSM with middle and large $\tan\beta$ (say, $>$ 8).
In the paper we shall perform a detailed analysis of $S_{\phi K_S}$
and $S_{\eta^\prime K}$ as well as Br in MSSM including NHB
contributions and the $\alpha_s$ corrections of hadronic matrix
elements.

We need to have new CP violation sources in addition to that of
CKM matrix in order to explain the deviations of $S_{\phi K_S}$ and
$S_{\eta^\prime K}$ from SM. There are new sources of flavor and
CP violation in MSSM. Besides the CKM matrix, the $6\times 6$
squark mass matrices are generally not diagonal in flavor
(generation) indices in the super-CKM basis in which superfields
are rotated in such a way that the mass matrices of the quark
field components of the superfields are diagonal. This rotation
non-alignment in the quark and squark sectors can induce large
flavor off-diagonal couplings such as the coupling of gluino to
the quark and squark which belong to different generations. These
couplings can be complex and consequently can induce CP violation
in flavor changing neutral currents (FCNC). It is well-known that
the effects of the primed counterparts of usual operators are
suppressed by ${m_s}/{m_b}$ and consequently negligible in SM
because they have the opposite chirality. However, in MSSM their
effects can be significant, since the flavor non-diagonal squark
mass matrix elements are free parameters which are only subjective
to constraints from experiments.

For the $b\to s$ transition, besides the SM contribution, there
are mainly two new contributions arising from the QCD and
chromomagnetic penguins and neutral Higgs boson (NHB) penguins
with the gluino and squark propagated in the loop in
MSSM\footnote{The chargino contributions have been studied in
Ref.\cite{cghk}}. The relevant Wilson coefficients at the $m_W$
scale have been calculated by using vertex mixing method in
Ref.\cite{hw}. There is the another method, "mass insertion
approximation"(MIA)~\cite{mi}, which works in flavor diagonal
gaugino couplings $\tilde{g}q\tilde{q}$ and diagonal quark mass
matrices with all the flavor changes rested on the off-diagonal
sfermion propagators. The MIA can be obtained in VM through Taylor
expansion of nearly degenerate squark masses $m_{\tilde{q}_i}$
around the common squark mass $m_{\tilde{q}}$, $m^2_{\tilde{q}_i}
\simeq m^2_{\tilde{q}} (1 + \Delta_i)$. Thus MIA can work well for
nearly degenerate squark masses and, in general, its reliability
can be checked only a posteriori. However, for its simplicity, it
has been widely used as a model independent analysis to find the
constraints on different off-diagonal parts of squark mass
matrices from experiments~\cite{gabbiani}. Therefore, we use MIA
to calculate Wilson coefficients in the paper.

As it is shown that both Br and CP asymmetries depend
significantly on how to calculate hadronic matrix elements of
local operators\cite{chw}. Recently, two groups, Li et al.
\cite{li1,li} and BBNS~\cite{bbns,bbns1}, have made  significant
progress in calculating hadronic matrix elements of local
operators relevant to charmless two-body nonleptonic decays of B
mesons in the PQCD framework. The key point to apply PQCD is to
prove that the factorization, the separation of the short-distance
dynamics and long-distance dynamics, can be performed for those
hadronic matrix elements. It has been shown that in the heavy
quark limit (i.e., $m_b\to\infty$) such a separation is indeed
valid and hadronic matrix elements can be expanded in $\alpha_s$
such that the tree level (i.e., the $\alpha_s^0$ order) is the
same as that in the naive factorization and the $\alpha_s$
corrections can be systematically calculated\cite{bbns}. Comparing
with the naive factorization, to include the $\alpha_s$ correction
decreases significantly the hadronic uncertainties. In particular,
the matrix elements of the chromomagnetic-dipole operators
$Q_{8g}^{(\prime)}$ have large uncertainties in the naive
factorization calculation which lead to the significant
uncertainty of the time dependent CP asymmetry in SUSY
models\cite{hlm}. The uncertainties are greatly decreased in BBNS
approach\cite{bbns1}. In the paper we shall use BBNS's approach (
QCD factorization ) to calculate the hadronic matrix element of
operators relevant to the decays $B\rightarrow \phi
K_S,\,\eta^\prime K_S$ up to the $\alpha_s$ order.

The experimental results, $S_{\phi K}$ is negative and
$S_{\eta^{\prime} K}$ is positive but smaller than 0.7, which implies that
new CP violating physics affects $B\to \phi K_S$ in a dramatic way
but gives $B\to \eta^{\prime} K_S$ a relatively small effect, which has been
thought to be problematic\cite{nir}. To solve the problem, Khalil
and Kou invoke to have operators with opposite chirality because
the decay constant of $\ep$ is sensitive to the chirality of the
quarks and that of $\phi$ is independent of the chirality of the
quarks, operators with opposite chirality give contributions with
opposite signs to $B\to\ep K_S$ but contributions with the same
sign to $B\to \phi K_S$\cite{kkou}. In MSSM the $LL$ ($LR$) and
$RR$ ($RL$) mass insertions give contributions to operators with
opposite chirality respectively. It is shown in Ref.\cite{kkou}
that (1) it is possible to explain the experimental results if
having both the $LL$ and $RR$ (or $LR$ and $RL$) mass insertions
simultaneously, and (2) it is impossible to explain the
experimental results if having only the LL and/or LR (or RR and/or
RL) insertions. The first claim is certainly valid in general.
However, the second claim is of the consequence of using the naive
factorization to calculate hadronic matrix elements of operators.
The current-current operators contribute to $B\to\ep K_S$ but not
to $B\to \phi K_S$. Their effects are not significant in both the
naive factorization approach and the QCD factorization approach.
However, the $\alpha_s$ corrections of hadronic matrix elements of
QCD penguin operators and chromomagnetic-dipole operators have
significant effects and consequently make the claim (2) in
Ref.\cite{kkou} not valid. We show in the paper that one can
explain the experimental results, $S_{\phi K}$ is negative and
$S_{\eta^\prime K}$ is positive but smaller than 0.7, with a
flavor non-diagonal mass insertion of any chirality when
$\alpha_s$ corrections of hadronic matrix elements of operators
are included. We show that in the case of $LL$ and $RR$ mass
insertions the Higgs mediated contributions to $S_{\phi K}$ alone
can provide a significant
deviation from the SM in some region of parameters and 
a possible explanation of $S_{\ep K}$ in $1 \sigma$ experimental
bounds in a very small region of parameters, satisfying all the
relevant experimental constraints. When including all the SUSY
contributions, there are regions of parameters larger than those
in the case of including only the Higgs mediated contributions in
which the theoretical predictions for $S_{\phi K}$ and $S_{\ep K}$
are in agreement with the data in $1 \sigma$ experimental bounds.
In the case of $LR$ and $RL$ insertions a satisfied explanation
can also be obtained. We show that the branching ratio of $B\to
\eta^{\prime} K_S$, which in SM with the naive factorization is
smaller than the measured value, can be in agreement with data due
to SUSY contributions in quite a large regions of parameters.

The paper is organised as follows. In section II  we  give the
effective Hamiltonian responsible for the $b\rightarrow s$
transition in MSSM. We give the Wilson coefficients of operators
including those induced by NHBs at LO in MIA with double
insertions. In Section III we present the decay amplitudes and the
CP asymmetries $S_{\phi K}$ and $S_{\ep K}$. In particular, the
hadronic matrix elements of NHB induced operators to the
$\alpha_s$ order are calculated. The Section IV is devoted to
numerical results. We draw conclusions and discussions in Section
V. The loop functions and some coefficients of the matrix element
of the effective Hamiltonian as well as implementation of
$\eta$--$\eta'$ mixing are given in Appendices.

\section{Effective Hamiltonian}
The effective Hamiltonian for $b \rightarrow s$ transition can be
expressed as\cite{bur,chw}
\begin{eqnarray}\label{eff}
 {\cal H}_{\rm eff} &=& \frac{G_F}{\sqrt2} \sum_{p=u,c} \!
   V_{pb} V^*_{ps} \bigg(C_1\,Q_1^p + C_2\,Q_2^p
   + \!\sum_{i=3,\dots, 16}\![ C_i\,Q_i+ C_i^\prime\,Q_i^\prime]
   \nonumber \\&& + C_{7\gamma}\,Q_{7\gamma}
   + C_{8g}\,Q_{8g}
   + C_{7\gamma}^\prime\,Q_{7\gamma}^\prime
   + C_{8g}^\prime \,Q_{8g}^\prime \, \bigg) + \mbox{h.c.} \,
\end{eqnarray}
Here $Q_i$ are quark and gluon operators and are given by
\footnote{For the operators in SM we use the conventions in
Ref.\cite{bbns1} where $Q_1$ and $Q_2$ are exchanged each other
with respect to the convention in most of papers.}
\begin{eqnarray}
&&Q_1^p = (\bar s_\alpha p_\beta)_{V-A} (\bar p_\beta
b_\alpha)_{V-A},\hspace{2.3cm}
Q_2^p = (\bar s_\alpha p_\alpha)_{V-A} (\bar p_\beta b_\beta)_{V-A},\nonumber\\
&&Q_{3(5)} = (\bar s_\alpha b_\alpha)_{V-A}\sum_{q} (\bar q_\beta
q_\beta)_{V-(+)A},\hspace{1cm} Q_{4(6)} = (\bar s_\alpha
b_\beta)_{V-A}\sum_{q}
(\bar q_\beta q_\alpha)_{V-(+)A},\nonumber\\
&&Q_{7(9)} = {3\over 2}(\bar s_\alpha b_\alpha)_{V-A}\sum_{q}
e_{q}(\bar q_\beta q_\beta)_{V+(-)A},\hspace{0.4cm}Q_{8(10)} ={3\over 2}
(\bar s_\alpha b_\beta)_{V-A}\sum_{q}
e_{q}(\bar q_\beta q_\alpha)_{V+(-)A},\nonumber\\
&&Q_{11(13)} = (\bar s\, b)_{S+P} \sum_q\,{m_q\over m_b} (\bar q\,
q)_{S-(+)P}\,,\nnb\\&&  Q_{12(14)} = (\bar s_i \,b_j)_{S+P}
 \sum_q\,{m_q\over m_b}(\bar q_j \,q_i)_{S-(+)P} \,, \nonumber\\
&&Q_{15} = \bar s \,\sigma^{\mu\nu}(1+\gamma_5) \,b
\sum_q\,{m_q\over m_b}
    \bar q\, \sigma_{\mu\nu}(1+\gamma_5)\,q \,,\nnb\\&&
Q_{16} = \bar s_i \,\sigma^{\mu\nu}(1+\gamma_5) \,b_j \sum_q\,
    {m_q\over m_b} \bar q_j\, \sigma_{\mu\nu}(1+\gamma_5) \,q_i
    \, ,\nnb\\
&&Q_{7\gamma} = {e\over 8\pi^2} m_b \bar s_\alpha \sigma^{\mu\nu}
F_{\mu\nu}
(1+\gamma_5)b_\beta, \nonumber\\
&&Q_{8g} = {g_s\over 8\pi^2} m_b \bar s_\alpha \sigma^{\mu\nu}
G_{\mu\nu}^a {\lambda_a^{\alpha \beta}\over 2}(1+\gamma_5)b_\beta,
\end{eqnarray}
where $(\bar q_1 q_2)_{V\pm A} =\bar q_1\gamma^\mu(1\pm\gamma_5)
q_2$, $(\bar q_1 q_2)_{S\pm P}=\bar q_1(1\pm\gamma_5)q_2$
\footnote{Strictly speaking, the sum over q in expressions of
$Q_i$ (i=11,...,16) should be separated into two parts: one is for
q=u, c, i.e., upper type quarks, the other for q=d, s, b, i.e.,
down type quarks, because the couplings of upper type quarks to
NHBs are different from those of down type quarks. In the case of
large $\tan\beta$ the former is suppressed by $\tan^{-1}\beta$
with respect to the latter and consequently can be neglected.
Hereafter we use, e.g., $C_{11}^c$ to denote the Wilson
coefficient of the operator $Q_{11}= (\bar s\, b)_{S+P}
\,{m_c\over m_b} (\bar c\, c)_{S-P}$.}, $p=u, c$, $q = u,d,s,c,b$,
$e_{q}$ is the electric charge number of $q$ quark, $\lambda_a$ is
the color SU(3) Gell-Mann matrix, $\alpha$ and $\beta$ are color
indices, and $F_{\mu\nu}$ ($G_{\mu\nu}$) are the photon (gluon)
fields strength.

The primed operators, the counterpart of the unprimed operators,
are obtained by replacing the chirality in the corresponding
unprimed operators with opposite ones. We calculate the SUSY
contributions due to gluino box and penguin diagrams to the
relevant Wilson coefficients at the $m_W$ scale in MIA with double
insertions, as investigated in Ref.~\cite{everett01}, which are
non-negligible if the mixing between left-handed and right-handed
sbottoms  is large and results are
\begin{eqnarray}\label{wilson}
C_3^{(\prime)} &=& - \frac{\alpha_s^2}{2\sqrt{2} G_F \lambda_t
m_{\tilde{g}}^2}
 [( - \frac{1}{9}b_1(x) - \frac{5}{9}b_2(x) -
 \frac{1}{18}p_1(x) - \frac{1}{2}p_2(x) )\delta^{dLL(RR)}_{23} + \nonumber\\
&& ( - \frac{1}{9}b^\prime_1(x) - \frac{5}{9}b^\prime_2(x) -
 \frac{1}{18}p^\prime_1(x) - \frac{1}{2}p^\prime_2(x) )
 \delta^{dLR(RL)}_{23}\delta^{dLR\ast(LR)}_{33} ] \nonumber\\
C_4^{(\prime)} &=& - \frac{\alpha_s^2}{2\sqrt{2} G_F \lambda_t
m_{\tilde{g}}^2}
 [( - \frac{7}{3}b_1(x) + \frac{1}{3}b_2(x) +
 \frac{1}{6}p_1(x) + \frac{3}{2}p_2(x) )\delta^{dLL(RR)}_{23} + \nonumber\\
&& ( - \frac{7}{3}b^\prime_1(x) + \frac{1}{3}b^\prime_2(x) +
 \frac{1}{6}p^\prime_1(x) + \frac{3}{2}p^\prime_2(x) )
 \delta^{dLR(RL)}_{23}\delta^{dLR\ast(LR)}_{33} ] \nonumber\\
C_5^{(\prime)} &=& - \frac{\alpha_s^2}{2\sqrt{2} G_F \lambda_t
m_{\tilde{g}}^2}
 [( \frac{10}{9}b_1(x) + \frac{1}{18}b_2(x) -
 \frac{1}{18}p_1(x) - \frac{1}{2}p_2(x) )\delta^{dLL(RR)}_{23} + \nonumber\\
&& ( \frac{10}{9}b^\prime_1(x) + \frac{1}{18}b^\prime_2(x) -
 \frac{1}{18}p^\prime_1(x) - \frac{1}{2}p^\prime_2(x) )
 \delta^{dLR(RL)}_{23}\delta^{dLR\ast(LR)}_{33} ] \nonumber\\
C_6^{(\prime)} &=& - \frac{\alpha_s^2}{2\sqrt{2} G_F \lambda_t
m_{\tilde{g}}^2}
 [( - \frac{2}{3}b_1(x) + \frac{7}{6}b_2(x) +
 \frac{1}{6}p_1(x) + \frac{3}{2}p_2(x) )\delta^{dLL(RR)}_{23} + \nonumber\\
&& ( - \frac{2}{3}b^\prime_1(x) + \frac{7}{6}b^\prime_2(x) +
 \frac{1}{6}p^\prime_1(x) + \frac{3}{2}p^\prime_2(x) )
 \delta^{dLR(RL)}_{23}\delta^{dLR\ast(LR)}_{33} ] \nonumber\\
C_{7\gamma}^{(\prime)} &=& - \frac{1}{27\lambda_t}
\frac{g_s^2}{g^2}
 \frac{m_w^2}{m_{\tilde{g}}^2} [
 F_2(x) \delta^{dLL(RR)}_{23} +
 F^\prime_2(x) \delta^{dLR(RL)}_{23} \delta^{dLR\ast(LR)}_{33} \nonumber\\
&& - 4 \frac{m_{\tilde{g}}}{m_b} F_4(x) \delta^{dLR(RL)}_{23}
 - 4 \frac{m_{\tilde{g}}}{m_b} F^\prime_4(x) \delta^{dLL(RR)}_{23}
 \delta^{dLR(LR\ast)}_{33} ] \nonumber\\
C_{8g}^{(\prime)} &=& - \frac{1}{72\lambda_t} \frac{g_s^2}{g^2}
 \frac{m_w^2}{m_{\tilde{g}}^2} [
 F_{12}(x) \delta^{dLL(RR)}_{23} +
 F^\prime_{12}(x) \delta^{dLR(RL)}_{23} \delta^{dLR\ast(LR)}_{33} \nonumber\\
&& - 8 \frac{m_{\tilde{g}}}{m_b} F_{34}(x) \delta^{dLR(RL)}_{23}
 - 8 \frac{m_{\tilde{g}}}{m_b} F^\prime_{34}(x) \delta^{dLL(RR)}_{23}
 \delta^{dLR(LR\ast)}_{33} ] \nonumber\\
C_{11}^{(\prime)} &=& \frac{e^2}{16\pi^2} \frac{m_b}{m_l}
 [C_{Q_1}^{(\prime)} \mp C_{Q_2}^{(\prime)}] \nonumber\\
C_{13}^{(\prime)} &=& \frac{e^2}{16\pi^2} \frac{m_b}{m_l}
 [C_{Q_1}^{(\prime)} \pm C_{Q_2}^{(\prime)}],\nnb\\
 C_i^{(\prime)} &=& 0,\,\,\,\, i=12,14,15,16
\end{eqnarray}
with $C_{Q_{1,2}}^{(\prime)}$\footnote{The operator of
$Q^{(\prime)}_{1,2}$ is defined as
$Q_1 = \frac{e^2}{8\pi^2} [\bar{s} (1 + \gamma_5) b][\bar{l} l]$,
$Q_1^\prime = \frac{e^2}{8\pi^2} [\bar{s} (1 - \gamma_5) b][\bar{l} l]$,
$Q_2 = \frac{e^2}{8\pi^2} [\bar{s} (1 + \gamma_5) b][\bar{l} \gamma_5 l]$,
$Q_2^\prime = \frac{e^2}{8\pi^2} [\bar{s} (1 - \gamma_5) b][\bar{l}
\gamma_5 l]$ } as
\begin{eqnarray}
C_{Q_1}^{(\prime)} &=& \frac{4}{3\lambda_t} \frac{g_s^2}{g^2
s_w^2}
 \frac{m_b m_l}{m_H^2} \frac{c_\alpha^2 + r_s s_\alpha^2}{c_\beta^2}
 \frac{m_{\tilde{g}}}{m_b} f^\prime_b(x) \delta^{dLL(RR)}_{23}
 \delta^{dLR(LR\ast)}_{33} \nonumber\\
C_{Q_2}^{(\prime)} &=& \mp \frac{4}{3\lambda_t} \frac{g_s^2}{g^2
s_w^2}
 \frac{m_b m_l}{m_A^2} (r_p + \tan^2\beta)
 \frac{m_{\tilde{g}}}{m_b} f^\prime_b(x) \delta^{dLL(RR)}_{23}
 \delta^{dLR(LR\ast)}_{33}
\end{eqnarray}
where $r_s = \frac{m^2_{H^0}}{m^2_{h^0}}$, $r_p =
\frac{m^2_{A^0}}{m^2_{Z^0}}$, and $x =
m_{\tilde{q}}^2/m_{\tilde{g}}^2$ with $m_{\tilde{q}}$ and
$m_{\tilde{g}}$ being the common squark mass and gluino mass
respectively. The one-loop functions in Eq. (\ref{wilson}) are
given in Appendix A. The parts of Wilson coefficients
$C_i^{(\prime)}$ (i= 3,...,6,$7\gamma$, 8g) which are obtained by
single insertion are the same as those in Ref.\cite{wil}. Differed
from the single insertion results, the LR or RL insertion also
generates the QCD penguin operators when one includes the double
insertions.

For the processes we are interested in this paper, the Wilson
coefficients should run to the scale of $O(m_b)$. $C_1-C_{10}$ are
expanded to $O(\alpha_s)$ and NLO renormalization group equations
(RGEs) should be used. However for the $C_{8g}$ and $C_{7\gamma}$,
LO results should be sufficient. The details of the running of
these Wilson coefficients can be found in Ref. \cite{bur}. The one
loop anomalous dimension matrices of the NHB induced operators can
be divided into two distangled groups~\cite{adm}
\begin{eqnarray}
\gamma^{(RL)}=\begin{tabular}{c|cccc} &$Q_{11}$&$Q_{12}$\\\hline
$Q_{11}$&$-16$&0\\
$Q_{12}$&-6&$2$
\end{tabular}
\end{eqnarray}
and
\begin{eqnarray}
\gamma^{(RR)}=\begin{tabular}{c|cccc}
&$Q_{13}$&$Q_{14}$&$Q_{15}$&$Q_{16}$\\\hline
$Q_{13}$&$-16$&0&1/3&$-1$\\
$Q_{14}$&$-6$&$2$&$-1$/2&$-7$/6\\
$Q_{15}$&16&$-4$8&$16/3$&0\\
$Q_{16}$&$-24$&$-56$&6&$-38/3$
\end{tabular}
\end{eqnarray}
Here and hereafter the factor, $\frac{\alpha_s}{4\pi}$, is
suppressed (i.e., the anomalous dimension matrix for $Q_{11,12}$
is $\frac{\alpha_s}{4\pi}\,\,\gamma^{(RL)}$, etc.). For
$Q_i^\prime$ operators we have
\begin{eqnarray}
\gamma^{(LR)}=\gamma^{(RL)}~~~~~{\rm and}~~~~~
\gamma^{(LL)}=\gamma^{(RR)}\,.
\end{eqnarray}
Because at present no NLO Wilson coefficients $C_i^{(\prime)}$,
i=11,...,16, are available we use the LO running of them in the
paper.

There is the mixing of the new operators induced by NHBs with the
operators in SM. The leading order anomalous dimensions have been
given in Refs.\cite{bghw,hk}. We list those relevant to our
calculations in the following. Defining \begin{eqnarray} \label{oq}
O_i=\frac{g^2}{16\pi^2}Q_{12+i},\,\,\,\,i=1,2,3,4,\end{eqnarray} one has
\begin{eqnarray} \gamma^{(RD)}=\begin{tabular}{c|cccc}
&$Q_{7\gamma}$&$Q_{8g}$\\\hline
$O_{1}$&$-1/3$&1\\
$O_{2}$&$-1$&$0$\\
$O_{3}$&28/3&$-4$\\
$O_{4}$&20/3&$-8$
\end{tabular}\label{q8g}
\end{eqnarray} The mixing of $Q_{11,12}$ onto the QCD penguin operators is
\begin{eqnarray} \gamma^{(MQ)}=\begin{tabular}{c|cccc}
&$Q_{3}$&$Q_{4}$&$Q_5$&$Q_6$\\\hline
$O_{11}$&$1/9$&$-1/3$&1/9&$-1/3$\\
$O_{12}$&0&$0$&0&0
\end{tabular}
\end{eqnarray} For the mixing among the primed operators, we have
\begin{eqnarray}
\gamma^{(LD^\prime)}=\gamma^{(RD)}~~~~~{\rm and}~~~~~ \gamma^{(M^\prime
Q^\prime)}=\gamma^{(MQ)}\,.
\end{eqnarray}
The mixing of the new operators induced by NHBs with the operators
in SM has non-negligible effects on the Wilson coefficients of the
SM operators at the $O(m_b)$ scale. In particular, the Wilson
coefficient of the chromo-magnetic dipole operator $C_{8g}$ at the
$O(m_b)$ scale, which has a large effect to $S_{M K}$ ($M=\phi,
\eta^\prime$), can significantly enhance due to the mixing. To see
it explicitly we concentrate on the mixing of $O_i$ (for its
definition, see Eq.(\ref{oq})) onto $Q_{8g}$. Solving RGEs, we
have
\begin{eqnarray}
C_{8g}(\mu)&=&\sum_{c=1,...,4}A(\mu_0)(\eta(\mu)^{\gamma_{cc}/2\beta_0}-
\eta(\mu)^{\gamma_{8g8g}/2\beta_0})+
C_{8g}(\mu_0)\eta^{\gamma_{8g8g}
/2\beta_0},\label{c8g}\\
A(\mu_0)&=&\sum_{a,b=1,...,4}\gamma_{a1}
V^{-1}_{ac}V_{cb}C_b(\mu_0)/(\gamma_{cc}
-\gamma_{8g8g})\\
\eta&=&\alpha_s(\mu_0)/\alpha_s(\mu), \end{eqnarray} where $V$ and
$\gamma_{aa}$ are given by \begin{eqnarray} V(\gamma^{(RR)}+
2\beta_0\,I)\,V^{-1}={\rm diag}(\gamma_{11},\gamma_{22},
\gamma_{33},\gamma_{44}).\end{eqnarray} with $I$ being the $4\times 4$ unit
matrix. Using \begin{eqnarray} C_a(\mu_0)&=& C_1(\mu_0)\delta_{a1}\end{eqnarray} and Eq.
(\ref{q8g}), Eq. (\ref{c8g}) reduces to
\begin{eqnarray} C_{8g}(\mu)=0.68
C_{8g}(\mu_0) - 3.2 C_{13}(\mu_0),\end{eqnarray}
where $C_1(\mu_0)=\frac{4\pi}{\alpha_s}C_{13}(\mu_0)$ has been used.

 In our numerical calculations we
neglect the contributions of \rm EW penguin operators $Q_{7,...10}$
since they are small compared with those of other operators.

\section{The decay amplitude and CP asymmetry}
 We use the BBNS approach~\cite{bbns,bbns1} to calculate the hadronic
 matrix elements of operators. In the approach the hadronic matrix
 element of a operator in the heavy quark limit can be written as
\begin{eqnarray}
\langle M K_S|Q|B\rangle  = \langle M K_S|Q|B\rangle _{f} [1+ \sum r_n \alpha_s^n ],
\end{eqnarray}
where $\langle M K_S|Q|B\rangle _{f}$ indicates the naive
factorization result and $M=\phi,\,\eta^\prime$. The second term
in the square bracket indicates higher order $\alpha_s$
corrections to the matrix elements~\cite{bbns1}. We calculate the
hadronic matrix elements to the $\alpha_s$ order in the paper. In
order to see explicitly the effects of new operators in the MSSM
we divide the decay amplitude into three parts. One has the same
form as that in SM, the second is for primed counterparts of the
SM operators, and the third is new which comes from the
contributions of Higgs penguin induced operators. That is, we can
write the decay amplitude for $B\to M K_S$ as 
\begin{eqnarray}
&&A(B\to M K_S) = {G_F\over \sqrt{2}} A \nnb\\
&& A= A^o+A^{o^\prime}+A^n, \\
&&
 A^{o^\prime}=\left\{\begin{array}{ll}
A^o(C_i\to C_i^\prime)  & \mbox{for\,$B\to \phi K_S$}\\
A^o(C_i\to -C_i^\prime) & \mbox{for\,$B\to \eta^\prime
K_S$}\end{array} \right.  \label{ap}
\end{eqnarray}
$A^o$ and $A^n$ will be given in subsections.

The time-dependent $CP$-asymmetry $S_{M K}$ is defined by
\begin{equation}
a_{M K}(t) = -C_{M K} \cos(\Delta M_{B_d^{0}} t) + S_{M K}
\sin(\Delta M_{B_d^{0}}t),
\end{equation}
where
\begin{equation}
C_{M K} = \frac{1-|\lambda_{M K}|^2}{1+|\lambda_{M K}|^2}
\,, \ \ \ \ \ \ \ \ \ \ S_{M K} = \frac{2\,\mathrm{Im}
\lambda_{M K}}{1+|\lambda_{M K}|^2} \; .\label{defi}
\end{equation}
Here $\lambda_{M K}$ is defined as
\begin{eqnarray}
\lambda_{M k} &=& \left(\frac{q}{p}\right)_B
\frac{\mathcal{A}(\overline{B} \rightarrow M K_S)}{\mathcal{A}(B
\rightarrow M K_S)}.
\end{eqnarray}
The ratio $(q/p)_B$
 is nearly a pure phase. In SM $\lambda_{M K}= e^{i2\beta}+ \! O(\lambda^2
 \!)$. As pointed out in Introduction, the MSSM can give a
 phase to the decay which we call
$\phi^{\mathrm{SUSY}}$. Then we have
\begin{equation}
\lambda = e^{i(2\beta+\phi^{\mathrm{SUSY}})}
\frac{|\bar{\mathcal{A}}|}{|\mathcal{A}|} \ \ \ \Rightarrow S_{M
K} = \sin (2\beta+\phi^{\mathrm{SUSY}})
\end{equation}
if the ratio $\frac{|\bar{\mathcal{A}}|}{|\mathcal{A}|}=1$. In
general the ratio in the MSSM is not equal to one and consequently
it has an effect on the value of $S_{M K}$, as can be seen from
Eq.(\ref{defi}). Thus the presence of the phases in the squark
mass matrix can alter the value of $S_{M K}$ from the standard
model prediction of $S_{M K}=\sin 2\beta_{J/\psi K} \sim 0.7$.

\subsection{$B_{d}^{0} \rightarrow \phi K_S$}

 $A^o$ for $B_{d}^{0} \rightarrow \phi K_S$, to the $\alpha_s$ order,
in the heavy quark limit is given as\cite{hz,hmw}\begin{eqnarray}
A^o&=&\langle \phi|\bar s\gamma_\mu s |0\rangle  \langle K|\bar s
\gamma^\mu b|B\rangle \times\sum_{p=u,c}  V_{pb} V^*_{ps}\left[
a_3 +a_4^p+a_5 - {1\over 2}(a_7 + a_9 + a_{10}^p )
 \right], \label{ao}\eea where $a_i$'s have been given in
Refs.\cite{hz,hmw} and are listed in Appendix B. The hadronic
matrix element of the vector current can be parameterized as
$\langle K|\bar s \gamma^\mu b|B\rangle  = F_1^{B\to K}(q^2)
(p_B^\mu + p_K^\mu) +(F^{B\to K}_0(q^2)-F^{B\to K}_1(q^2))
(m_B^2-m_K^2)q^\mu/q^2$. For the matrix element of the vector
current between the vacuum and $\phi$, we have $\langle \phi| \bar
s \gamma_\mu b | 0 \rangle = m_\phi f_\phi \epsilon^\phi_\mu$.

 $A^n$ for $B_{d}^{0}
\rightarrow \phi K_S$, to the $\alpha_s$ order, in the heavy quark
limit is given as\cite{chw} \bea
A^n&=&A^n(C_i)+A^n(C_i\rightarrow C_i^\prime),\nnb\\
A^n(C_i)&=&\langle \phi|\bar s\gamma_\mu s |0\rangle \langle
K|\bar s \gamma^\mu b|B\rangle \,(- V_{tb} V^*_{ts}) \left[
a_4^{neu}+ {m_s \over m_b}\left( -{1\over 2}a_{12}+ {4 m_s\over
m_b}\,a_{15} \right) \right] \,.\label{an}\end{eqnarray} $a_i$'s
in Eq. (\ref{an})  are
\begin{eqnarray}\label{aiphi}
\vspace{0.1cm}
   a_4^{neu} &=& {C_F\alpha_s\over 4\pi}\,{P_{\phi,2}^{neu}\over N_c}
\,,\nonumber\\
   a_{12} &=& C_{12} + {C_{11}\over N_c} \left[ 1
    + {C_F \alpha _s \over 4 \pi  } \left( -V_{\phi}^\prime
     -{4 \pi  ^2 \over N_c} H_{K_S\phi } \right) \right]\,,\nonumber\\
   a_{15} &=&  C_{15} + \displaystyle{\frac{C_{16}}{N_c}
   }\,,\label{anew}
\end{eqnarray}
with $P_{\phi,2}^{neu} $ being
\begin{eqnarray}\label{P2phi}
   P_{\phi,2}^{neu} &=&  -{1\over 2}
 C_{11}  \times  \left[ {m_s \over m_b}  \left( {4\over 3} \ln {
m_b \over \mu } - G_\phi (0)\right)  +\left( { 4\over 3} \ln { m_b
\over \mu } - G_ \phi (1) \right) \right] \nonumber \\
  && + C_{13}  \left[ -2 \ln { m_b \over \mu }
       \,G^0_\phi - GF_ \phi (1) \right]
       - 4 C_{15}  \left[ \left( -{1\over 2}
-2 \ln { m_b \over \mu }\right) G^0_\phi - GF_\phi (1) \right] \nonumber\\
 &&{}- 8  C_{16}^c
 \bigg[ \left( {m_c \over m_b } \right)  ^2 \,
  \left( - 2 \ln { m_b \over \mu }\,G^0_\phi - GF_ \phi(s_c )
  \right)\bigg] - 8   C_{16} \left[ -2 \ln { m_b\over \mu }\,G^0_\phi
 - GF_\phi(1)
 \right]
\end{eqnarray}
where $s_c=m_c^2/m_b^2$ and
\begin{eqnarray}
&&G^0_\phi = \int_0^1 \frac{dx}{\bar x}\,
    \Phi_{\phi}(x)\, ,~~~~~~~GF(s,x) = \int^1_0 dt \ln \big[ s-x \,t {\bar t} \big]\, ,
   \nonumber \\
&& GF_\phi (s) = \int^1_0 dx { \Phi _\phi (x)\over \bar x} \,
GF(s-i\, \epsilon , \bar x)\,
\end{eqnarray}
with $\bar x =1-x$ and $\Phi_{\phi}(x)= 6 x \bar x$ in the
asymptotic limit of the leading-twist distribution amplitude. In
Eq. (\ref{anew}) the expressions of $V^\prime$ and $H_{K_S\phi }$
can be found in the Appendix B. In calculations we have set
$m_{u,d}=0$ and neglected the terms which are proportional to
$m_s^2/m_b^2$ in Eq.(\ref{P2phi}). We have
included only the leading twist contributions in Eq.(\ref{aiphi}). 
Here we have used the BBNS approach~\cite{bbns,bbns1} to calculate
the hadronic matrix elements of operators. 

\subsection{$B_{d}^{0} \rightarrow \eta^\prime K_S$}

$A^o$  for $B\to \ep K_S$, to the $\alpha_s$ order, in the heavy
quark limit has been given in Ref.\cite{bn} as
 \begin{eqnarray}
A^o&=&\sum_{p=u,c}V_{pb} V^*_{ps}\,m_B^2\, A^o_p \nnb\\
  A^o_p&=& F^{B\to K}\left\{ f^u_{\eta'} \left[
\delta_{up}a^p_2(K\eta') + 2( a_3(K\eta') - a_5(K\eta'))
-\mu^s_{\eta'}\left( a_6^p(K\eta')-a_8^p(K\eta')\right) + {1\over
2} (-a_7(K\eta') + a_9(K\eta') ) \right] \right.
\nonumber\\
&&\left.+ f^s_{\eta'} \left[ a_3(K\eta') + a_4^p(K\eta') -a_5(K\eta') +
\mu^s_{\eta' } \left( a_6^p(K\eta')-a_8^p(K\eta')
\right) - {1\over 2} ( -a_7(K\eta') +a_9(K\eta') )
- {1\over 2} a^p_{10}(K\eta') \right] \right\}
\nonumber\\
&& + F^{B\to \eta'} f_K \left[  a_4^p(\eta'K) +\mu_K \left(
a_6^p(\eta'K)-a_8^p(\eta'K) \right) - {1\over 2} a_{10}^p(\eta'K)
\right], \label{ao2} \eea where $F^{B\to M}$ (M=$\eta^\prime$, K)
is a $B\to M$ form factor calculated at $q^2=m_{K}^2$ or
$m_{\eta^\prime}^2 \approx 0$. The $a_i$'s in Eq. (\ref{ao2}) have
been given in Ref.\cite{bn} and are listed in Appendix B.

We calculate $A^n$ for $B\to \ep K_S$, to the $\alpha_s$ order, in
the heavy quark limit and the result is \bea \label{aetap}
A^n&=&-V_{tb} V_{ts}^*\,m_B^2\,\left[ A^n(C_i)+ A^n(C_i\rightarrow
-C_i^\prime)\right] ,\nnb\\
A^n(C_i)&=& F^{B\to K}\left\{ f^q_{\eta'} \left[
-\mu^s_{\eta'}\left( a_6^n(K\eta')-a_8^n(K\eta')\right) \right] +
f^s_{\eta'} \left[ a_4^n(K\eta')  + \mu^s_{\eta' } \left(
a_6^n(K\eta')-a_8^n(K\eta') \right)  - {1\over 2} a^n_{10}(K\eta')
\right] \right\}
\nonumber\\
&& + F^{B\to \eta'} f_K \left[  a_4^n(\eta'K) +\mu_K \left( a_6^n(\eta'K)-a_8^n(\eta'K) \right) -
{1\over 2} a_{10}^n(\eta'K) \right]\nonumber\\
&&{}+ F^{B\to K } \left\{{m_d\over m_b} f^q_{\eta'}\left[{1\over
2} \mu^s_{\eta'}\left( -a_{11} + a_{13} - {1\over 2} a_{14}
\right) \right] + {m_s\over m_b}f^s_{\eta'} \left[ - {1\over 2}
a_{12} + {1\over 2} \mu^s_{\eta'}
\left( a_{11} - a_{13} + {1\over 2} a_{14} \right) \right] \right\} \nonumber\\
&&{}+ F^{B\to \eta'} {m_d\over m_b} f_K \left[ -{1\over 2} a_{12}
+\mu_K \left( {1\over 4} a_{14} -3 a_{16} \right) \right]
\,,\label{an1}\end{eqnarray} where decay constants
$f^q_{\eta^\prime}, f^s_{\eta^\prime}$ are defined in Appendix C,
$\mu_K=m_b~R_K/2$, $\mu^s_{\eta^\prime}=m_b~R^s_{\eta^\prime}$,
and $a_i^n$'s and $a_i$'s are defined as
\begin{eqnarray}\label{aeta}
a_4^n (M_1M_2) &=&{\alpha_s\over 4\pi}{C_F\over N_c}P_{M_2, 2}^n, \hspace{1.3cm}
a_6^n (M_1M_2) = {\alpha_s\over 4\pi}{C_F\over N_c}
P_{M_2, 3}^n,\nnb\\
a_8^n (M_1M_2) &=&{\alpha_{\rm em}\over
9\pi}{C_F\over N_c} P_{M_2, 3}^{n,\rm EW},\qquad
a_{10}^n (M_1M_2) ={\alpha_{\rm em}\over
9\pi}{C_F\over N_c} P_{M_2, 2}^{n,\rm EW}, \nnb \\
a_{11}(M_1M_2)&=& C_{11}+{C_{12}\over N_c }\nnb\\
   a_{12} (M_1M_2)&=& C_{12} + {C_{11}\over N_c} \left[ 1
    + {C_F \alpha _s \over 4 \pi  } \left( -V_{M_2}^\prime
     -{4 \pi  ^2 \over N_c} H_{M_1M_2 } \right) \right]\,,\nonumber\\
a_{13}(M_1M_2)&=& C_{13}+{C_{14}\over N_c }+C_{16}{\alpha_s \over
4\pi}{C_F\over N_c}\,(-8 V_{13})\nnb\\
a_{14} (M_1M_2)&=& C_{14}+{C_{13}\over N_c }+ 2 {\alpha_s \over
4\pi}{C_F\over N_c}\,(C_{13}V_{13}+ C_{15}V_{15})
\end{eqnarray}
where \bea V_{13}&=&\int_0^1 du\ \phi_{M_2}(u)V_{S,2}+\int_0^1 du
dv d\xi H_{S,3}\nnb\\ V_{15}&=&\int_0^1 du\ \phi_{M_2}(u)V_{T,2}-4
\int_0^1 du dv d\xi H_{S,3}\label{aeta1} \eea and $V'$,
$H_{M_1M_2}$ could be found in Appendix B. In
Eqs.(\ref{aeta},\ref{aeta1}) $V_{S,2}$ and $V_{T,2}$ come from
vertex contributions, $P$'s from penguin diagrams, and $H_{S,3}$
from hard-scattering contributions, which will be shown in the
following. In numerical calculations we set $m_{u,d}=0$ so that
the terms which are proportional $m_{q}$ (q=u, d) in Eq.
(\ref{aetap}) are neglected.
\begin{figure}[t]
   \epsfxsize=10cm
   \centerline{\epsffile{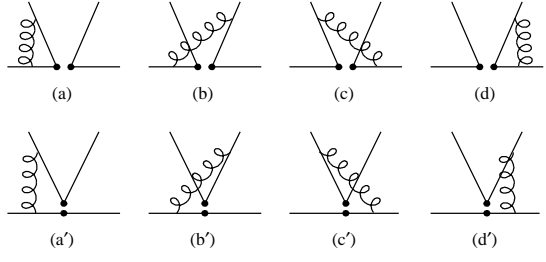}}
   \vspace*{-2.7cm}
\caption[dummy]{\label{vertex}vertex corrections}
\end{figure}
\begin{itemize}
\item{\bf Vertex Contributions}

The vertex contributions come from {\bf Fig}.(\ref{vertex}). In
the calculations of vertex corrections we need to distinguish two
diagrams with different topologies showed in {\bf
Fig}.(\ref{vertex}) when we insert different operators. In the
calculation of twist-3 contributions we find the infrared
divergence cancels only if we include all four diagrams for each
topology and use the asymptotic form of twist-3 distribution
amplitude, i.e., $\phi_p(x)=1, \phi_\sigma(x)=6x(1-x)$, which can
be combined into $-{i\over 4 }f_P {\gamma_5 \kslash_2\,
\kslash_1\, \over k_2\cdot k_1 }$\cite{gt,bbns1}. By a
straightforward calculation, we obtain
\begin{eqnarray}
 V_{S,2}&=&  \int^1_0 du \Bigg\{ \left[2
+2\ln u -2 {\bar u}  \left({\rm Li}_2 \left(1-{1\over u}\right) -{\rm
Li}_2 \left(1-{1\over {\bar u}}  \right) \right) - 2 {\bar u}
\,(\ln u-\ln {\bar u} )\pi i  \right]
+\left[ -11 - 6 \ln{\mu^2\over m_b ^2} + 3 \pi i
\right]  \Bigg\}\nonumber\\
 V_{T,2}&=&  \int^1_0 du \Bigg\{ -4 \left[ 2
+2\ln  u - 2 {\bar u}  \left({\rm Li}_2 \left(1- {1\over u }\right) -
{\rm Li}_2 \left(1- {1\over {\bar u}} \right) \right) -i\pi\  2 {\bar u}
(\ln u -\ln {\bar u} ) \right]
+\left[ 32 +24 \ln{\mu^2\over m_b ^2} -12 \pi i \right] \Bigg\}\nonumber\\
\end{eqnarray}
\begin{figure}[t]
\epsfxsize=8cm \centerline{\epsffile{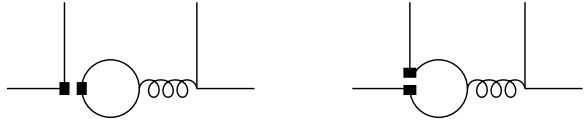}}
\centerline{\parbox{14cm}{\caption{\label{penguin} The two
different penguin contractions }}}
\end{figure}

\item{\bf Penguin Contributions}\\

In the calculations of penguin contributions we also need to
consider the difference between two diagrams with different
topologies showed in {\bf Fig}.(\ref{penguin}). We obtain that
the twist-2 parts of penguin contributions  $P_{M_2, 2}^n, P_{M_2,
2}^{\rm \rm EW}$ are
\begin{eqnarray}
P_{M_2, 2}^n &=&
  -{1\over 2}
  C_{11}
\times  \left[ {m_s \over m_b}  \left({4\over 3} \ln { m_b \over
\mu } - G(0)\right) +\left({ 4\over 3} \ln { m_b
\over \mu } - G (1) \right) \right] \nonumber \\
  &&{}+ C_{13} \left[ -2 \ln { m_b \over \mu }
       \,G^0_{M_2}- GF_{M_2} (1) \right]
       - 4  C_{15} \left[ \left(-{1\over 2}
-2 \ln { m_b \over \mu }\right) G^0_{M_2}- GF(1) \right] \nonumber\\
&& -8   C_{16} \left[-2 \ln { m_b\over \mu
}\,G^0_{M_2}
 - GF_{M_2}(1) \right]-8   C_{16}^c\left[ \left({m_c \over
m_b } \right)  ^2 \,
  \left(- 2 \ln { m_b \over \mu }\,G^0_{M_2}- GF_{M_2}(s_c) \right) \right] \\
   P_{M_2, 2}^{n,\rm EW}&=&-{1\over 2}\left\{
  C_{12}
\times  \left[ {m_s \over m_b}  \left({4\over 3} \ln { m_b \over
\mu } - G(0)\right) +\left({ 4\over 3} \ln { m_b
\over \mu } - G (1) \right) \right]\right. \nonumber \\
  &&{}+\left[  C_{13}+ N_c  C_{14}\right]
  \left[ -2 \ln { m_b \over \mu }
       \,G^0_{M_2}- GF_{M_2} (1) \right] \nonumber\\
&&{}-8\left[ N_c  C_{15}+ C_{16}\right]
  \left[ \left(-{1\over 2}
-2 \ln { m_b \over \mu }\right) G^0_{M_2}- GF_{M_2}(1) \right] \nonumber\\
&&{}-4 \left[ C_{15}+ N_c  C_{16}\right]
\left(-2 \ln { m_b\over \mu }\,G^0_{M_2}
 - GF_{M_2}(1) \right)  \nonumber\\
&&{}\left.-4 \left[ C_{15}^c+ N_c C_{16}^c
 \right]\left[\left({m_c \over m_b } \right)  ^2 \,
  \left(- 2 \ln { m_b \over \mu }\,G^0_{M_2}- GF_{M_2}(s_c)
\right)\right]\right\}
  \end{eqnarray}
where $s_c=(m_c/m_b)^2, G^0_{M_2}=\int^1_0 dx \phi_{M_2}(x)/x
$ and the function $GF_{M_2}(s)$ is given by
\begin{eqnarray}
 GF_{M_2}(s_c)&=& \int^1_0 dx\ GF(s_c, x){\phi_{M_2}(x)\over x}
\end{eqnarray}

  Compared with twist-2 contributions, the twist-3 projection yields an
  additional factor of ${\bar x}$ which has appeared in penguin contributions
  proportional to $C_{7\gamma}^{\rm eff}$ and $C_{8g}^{\rm
  eff}$.\citation{beneke01} We therefore find
  \begin{eqnarray}
P_{M_2, 3}^n &=& -{1\over 2}  C_{11}
\times  \left[ {m_s \over m_b}  \left({4\over 3} \ln { m_b \over
\mu } - \widehat{G}(0)\right) +\left({ 4\over 3} \ln { m_b
\over \mu } - \widehat{G} (1) \right) \right] \nonumber \\
  &&{}+ C_{13} \left[ -2 \ln { m_b \over \mu }
       - \widehat{GF} (1) \right]
       - 4  C_{15} \left[ \left(-{1\over 2}
-2 \ln { m_b \over \mu }\right) - \widehat{GF}(1) \right] \nonumber\\
&& -8   C_{16} \left[-2 \ln { m_b\over \mu }
 - \widehat{GF}(1) \right]-8
  C_{16}^c\left[ \left({m_c \over
m_b } \right)  ^2 \,
  \left(- 2 \ln { m_b \over \mu }- \widehat{GF}(s_c) \right) \right] \\
P_{M_2, 3}^{n,\rm EW}&=&-{1\over 2}\left\{ C_{12}
\times  \left[ {m_s \over m_b}  \left({4\over 3} \ln { m_b \over
\mu } - \widehat{G}(0)\right) +\left({ 4\over 3} \ln { m_b
\over \mu } - \widehat{G} (1) \right) \right]\right. \nonumber \\
  &&{}+\left[  C_{13}+ N_c  C_{14}\right]
  \left[ -2 \ln { m_b \over \mu }
       - \widehat{GF} (1) \right] -8\left[ N_c  C_{15}+ C_{16}\right]
  \left[ \left(-{1\over 2}
-2 \ln { m_b \over \mu }\right) - \widehat{GF}(1) \right] \nonumber\\
&& \left.-4 \left[ C_{15}+ N_c  C_{16}\right]
\left(-2 \ln { m_b\over \mu }
 - \widehat{GF}(1) \right)  -4 \left[ C_{15}^c+ N_c C_{16}^c
 \right]\left[\left({m_c \over m_b } \right)  ^2 \,
\left(- 2 \ln { m_b \over \mu }
- \widehat{GF}(s_c) \right)\right]\right\}
\end{eqnarray} with \begin{eqnarray}
  \widehat{GF}(s_c)&=& \int^1_0 dx\ GF(s_c, x)\,{\phi_{M_2}(x)}.
\end{eqnarray}

\begin{figure}[t]
   \vspace*{-1.7cm}
   \epsfxsize=10cm
   \epsfysize=8cm
   \centerline{\epsffile{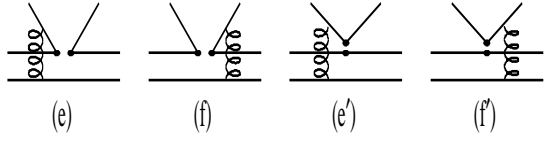}}
   \vspace*{-1.7cm}
\caption[dummy]{\label{hardscattering}hard-scattering contributions}
\end{figure}
\item{\bf Hard-scattering Contributions}\\
Hard-scattering contributions come from {\bf Fig.}(\ref{hardscattering}) which
read
 \begin{eqnarray} H_{S,3}&=& {1\over m_B^2 F^{B\to M_1}_1}
{1 \over 8u^2 \xi v } \left[ - u  \, \mu_{M_1}  \,
\phi_B(\xi) \, {\phi_{{M_1}\sigma}'(u)\over 6} \, {\phi_{{M_2}\sigma}'
(v)\over 6}
+3 v m_B \, \phi_B(\xi) \, \phi_{M_1}(u) \, \phi_{M_2p}(v) \right.\nonumber\\
&&- 3(u+v)  \, \mu_{M_1}  \phi_B(\xi)
\phi_{M_1p}(u) \phi_{M_2p}(v)   -(u-v)  \,\mu_{M_1}  \, \phi_B(\xi)
{\phi_{M_1\sigma}'(u)\over 6} \phi_{M_2p}(v) \nonumber\\
&&\left.+u  \,\mu_{M_1}  \,
\phi_B(\xi) \phi_{M_1p}(u) {\phi_{M_2\sigma}'(v)\over 6} \right]\label{hs3},
\end{eqnarray}
where $\phi'$ means the derivative of $\phi$. There is end-point
singularity in Eq.(\ref{hs3}) which can
be treated in the way given in Appendix B.
\begin{figure}
\epsfxsize=3.5cm
\centerline{\epsffile{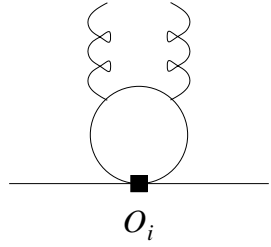}}
\vspace{0.0cm}
\centerline{\parbox{14cm}{\caption{\label{fig:2gluon_kernel}
Two-gluon emission from a quark loop. A second diagram with the two
gluons crossed is implied.}}}
\end{figure}

Because $\eta^\prime$ is a flavor singlet there are three extra
contributions related to the gluon content of $\eta^\prime$. They
come from: the $b\to s g g$ amplitude, spectator scattering
involving two gluons, and singlet weak annihilation. As analysed
in Ref.\cite{bn}, the singlet weak annihilation is suppressed by
at least one power of $\Lambda_{QCD}/m_b$ in the heavy quark
limit. So we need to calculate the contributions from the $b\to s
g g$ decay and spectator scattering which have not included in the
results given above. In SM the relevant calculations have been
carried out and results have been given in Ref.\cite{bn}. We
calculate them with insertions of new operators. In order to make
the paper self-contained we also list the SM results given in
Ref.~\cite{bn}.
\end{itemize}

{\bf Two gluon mechanism}\\

\begin{itemize}
\item {\bf The $b\to sgg $ amplitude}\\

The amplitude for $b\to s g g$ comes from {\bf Fig.}
(\ref{fig:2gluon_kernel}). We use the formula (16) in
Ref.~\cite{bn} directly. For an operator $O=({\bar s}\Gamma_1
b)({\bar q} \Gamma_2 q)$, where the $\Gamma$'s denote arbitrary
spinor and color matrices, one has (assuming the two gluons to be
in a color-singlet configuration)
\begin{equation}
   {\cal A}(b\to s g g)|_{{\rm {\bf Fig}.~\ref{fig:2gluon_kernel}}}
   = - (\bar u_s\Gamma_1 u_b)\,
   \langle g(q_1) g(q_2)|\text{tr}\,(\Gamma_2\,A)|0\rangle \,,
\end{equation}
where
\begin{equation}
   A = \frac{\alpha_s}{4\pi N_c}\,\bigg\{
   \frac{1}{12 m_q}\,G_{\mu\nu}^A\,G^{A,\mu\nu}
   - \frac{(\qslash-6 m_q)\,i\gamma_5}{48 m_q^2}\,
   G_{\mu\nu}^A\,\widetilde{G}^{A,\mu\nu} + O(1/m_q^3) \bigg\} \,.
\end{equation}
When $\Gamma=V\pm A$, we obtain the contribution to the ${\bar B}
\to {\bar K} P$ decay amplitudes
\begin{equation}\label{anomaly}
   {\cal A}_p^{\rm charm} = \frac{a_P}{12 m_c^2}\,
   \langle \bar K(p')|\bar s\qslash(1-\gamma_5)b|\bar B(p)\rangle
   \left\{ \left( C_2 + \frac{C_1}{N_c} \right)
   \delta_{pc}+ \left( C_3-C_5 + \frac{C_4-C_6}{N_c} \right)
   \right\} .
\end{equation}
When $\Gamma=S\pm P$, we obtain that the decay amplitudes of
${\bar B} \to {\bar K} P$ are given as
\begin{eqnarray}
{\cal A}_{p, n}^{\rm  charm} &=& {a_P\over 2 m_c }\,
   \langle \bar K(p')|\bar s (1+\gamma_5)b|\bar B(p)\rangle \left\{ \left(
   C_{11}+{C_{12}\over N_c} \right) -\left( C_{13}+{C_{14}\over N_c} \right)
   \right\}
   \end{eqnarray}
\begin{figure}
\epsfxsize=10.0cm
\centerline{\epsffile{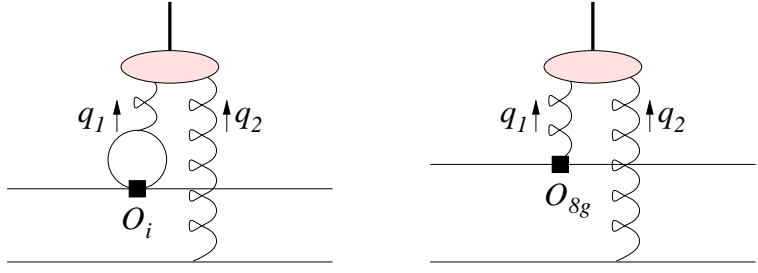}}
\vspace{0.2cm}
\centerline{\parbox{14cm}{\caption{\label{fig:2gluon_O8}
Spectator-scattering contributions to the $B\to K\eta^{\prime}$
decay amplitudes. The shaded blob represents the
$\eta^{(\prime)} g^* g^*$ form factor.}}}
\end{figure}

\item {\bf Spectator Mechanism}\\

The {\bf Fig.}(\ref{fig:2gluon_O8}) gives the spectator scattering
contributions. By a straightforward calculation, we obtain the
hard spectator-scattering contributions to the $B\to K\eta'$ decay
amplitude using two-gluon mechanism:
\begin{eqnarray} \label{hard}
A^{\rm spec} &&= {3 C_F \alpha_s ^2 \over N_c^2 } C_p f_B f_K \int^1_0
{\phi_B(\xi )\over \xi} \int^1_0 {dy \over y} \left[ P^p_2 \left( \phi_K(y) +
r^K_\chi \phi_p(y) \right) + P^{\rm neu}_2 \phi_K(y)+ r^K_\chi P^{\rm neu}_3
\phi_p(y) \right] \end{eqnarray}
with\begin{eqnarray}\label{P2y}
   P_2^p(y) &=& C_1 \left[ \frac43\ln\frac{m_b}{\mu} + \frac23
    - G(s_p,\bar y) \right]
    + C_3 \left[ \frac83\ln\frac{m_b}{\mu} + \frac43
    - G(0,\bar y) - G(1,\bar y) \right] \nonumber\\
   &&+ (C_4+C_6) \left[ \frac{4n_f}{3}\ln\frac{m_b}{\mu}
    - (n_f-2)\,G(0,\bar y) - G(s_c,\bar y) - G(1,\bar y) \right]\nonumber\\
    &&- {1\over 2} C_{11} \left[ {m_s\over m_b}  \left( {4\over 3 } \ln
      {m_b\mu} -G(0, \bar y)\right)  +\left( {4\over 3}\ln{m_b\over
      \mu}-G(1, \bar y)\right) \right] \\
P_2^{\rm neu}(y) &=& -2 C_{8g}^{\rm eff} +C_{13} \left[ -2 \ln {m_b\over \mu}
-GF(1, \bar y)\right] \nonumber\\
&&-4 C_{15} \left[ - {1\over 2} -2 \ln {m_b\over \mu}
-GF(1, \bar y) \right]-8C_{16} \left[ -2 \ln {m_b\over \mu} -GF(1, \bar
y)\right] \\
P_3^{\rm neu}(y) &=& -2 C_{8g}^{\rm eff} +C_{13} \left[ +1-2 \ln {m_b\over \mu}
-GF(1, \bar y)\right] \nonumber\\
&&-4 C_{15} \left[ - {3\over 2} -2 \ln {m_b\over \mu} -GF(1, \bar
y) \right]-8C_{16} \left[ -1-2 \ln {m_b\over \mu} -GF(1, \bar
y)\right]\end{eqnarray} with $s_q=(m_q/m_b)^2$, $n_f=5$. Obviously
there is singularity arising from the end-point region in the
Eq.(\ref{hard}) which makes us to consider the effect of $k_\perp$
in the end-point region. After involved the $k_\perp$ effect of
spectator quark and turned to $b$-space, the conjugate space of
$k_\perp$, Eq.(\ref{hard}) now reads
\begin{eqnarray} \label{hard1}
   {\cal A}_p^{\rm spec} &=&  \frac{3C_F\alpha_s^2}{4 N_c^2}\,C_P\,
   f_B f_K\, m_B^2 \int^\infty_0 bdb \int^1_0 d\xi dy \, {\mathscr P}_B(\xi, b)
   \nonumber\\
&&{}\times  \bigg\{ - \left[ K_0(b\sqrt{-c} ) - K_0(b\sqrt{a})\right]
\left[ P^p_2(y)\left( {\mathscr P}_K(y, b) + r^y_\chi {\mathscr P}_p(y, b)\right)
+r^y_\chi P^{\rm neu}_3 {\mathscr P}_p(y, b)  \right]\nonumber\\
&&\hspace{0.8cm} + m_B \left[ {b
K_1(b\sqrt{-c})\over 2 \sqrt{-{\bar y} }}- {K_0(b\sqrt{-c})-K_0 (b\sqrt{a})
\over {\bar y} m_B } \right] P^{neu}_2(y){\mathscr P}_K(y, b)  \bigg\}
\end{eqnarray}
where ${\mathscr P}_B(\xi, b), {\mathscr P}_K(y, b), {\mathscr
P}_p(y, b)$ are corrected distribution amplitude including Sudakov
factor~\cite{li,li1} and $K_{0(1)}$ are modified Bessel function of
order $0(1)$. Numerically we find that the contributions from the
two gluon mechanism are negligible.
\end{itemize}

\section{Numerical Results}
\subsection{Parameters input}

In our numerical calculations the following values are needed:
\begin{itemize}
\item{\bf Parameters related mixing of $\eta-\eta^\prime$} \\We
use the following input~\cite{Kaiser:1998ds}:
\begin{eqnarray}
\begin{array}{l@{\qquad}l@{\,}l}
  & f_q = f_\pi \,,
  & f_s = \sqrt{2 f_K^2-f_\pi^2} = 1.41 f_\pi \,, \\
  & h_q = f_q\,m_\pi^2 = 0.0025\,\text{GeV}^3 \,, \qquad
  & h_s = f_s\,(2m_K^2-m_\pi^2) = 0.086\,\text{GeV}^3 \,,\\
  & \phi=39.3^\circ\pm 1.0^\circ\,.
\end{array}
\end{eqnarray} \item{\bf Lifetime, mass and decay constants}
\begin{eqnarray}
\begin{array}{ccccc}
&\tau(B^0)=1.56\times 10^{-12}s,&
 M_B = 5.28 { { \rm GeV}}, &m_b =4.2{ \rm GeV},&\\
 &m_c =1.3 { \rm GeV},& m_s=100{  \rm MeV},&f_B = 0.190{ \rm GeV}, &
f_K=0.158 { \rm GeV}.
\end{array}
\end{eqnarray}
\item{\bf Chiral enhancement factors}\\
\vspace{0.2cm} For the chiral enhancement factors for the
pseudoscalar mesons, $\mu_{P}=\frac{m_b R_{P}}{2}$, we take
$$
R_{K^{\pm,0}}=R_{\pi^{\pm}}\simeq 1.2,
$$
which are consistent with the values used in \cite{bbns1,chenghy},
and
$$
R^s_{\eta^{(\prime)}}=\frac{m^2_{\eta^{(\prime)}}}{m_s m_b}.
$$
\item{\bf Wolfenstein parameters}\\
We use the Wolfenstein parameters fitted by Ciuchini et al\cite{ciuchini}:
\begin{eqnarray}
\begin{array}{ll@{\,}@{\qquad}l}
&A=0.819\pm 0.040&\lambda=0.2237\pm 0.0033\mbox{,}
\\
&{\bar \rho}=\rho (1-\lambda^2/2) = 0.224\pm 0.038\mbox{,}&\quad \rho
= 0.230\pm 0.039\mbox{,}\\
& {\bar \eta}=\eta (1-\lambda^2/2)
= 0.317\pm 0.040\mbox{,}&\eta=0.325\pm 0.039\mbox{,}\\
&\gamma = (54.8\pm 6.2)^\circ\mbox{,}& \sqrt{\rho^2+\eta^2} = 0.398\pm
0.040\,.
\end{array}
\end{eqnarray}
\item{\bf Form factors}\\
In the paper we need two form factors: $F^{B\to K}(0)=0.34$ and $F^{B\to
\eta'}$.
Compared with these rather well studied form factors, the form factors
for $B\to \eta^{(\prime)}$ are poorly known, which has hindered theoretical
predictions for B decays involving $\eta^{(\prime)}$ very much for a long time.
 We adopt the following parameterization for the form factors ($P=\eta$
or $\eta'$)\cite{bn}:
\begin{equation}
   F_0^{B\to P}(0) = F_1\,\frac{f_P^q}{f_\pi}
   + F_2\,\frac{\sqrt2 f_P^q+f_P^s}{\sqrt3 f_\pi} \,,
\end{equation}
where $F_1$ and $F_2$ both scale like $(\Lambda/m_b)^{3/2}$ in the
heavy-quark limit. In our numerical analysis we set
$F_1=F_0^{B\to\pi}(0)=0.28$ and take $F_2=0$ .
\end{itemize}

\subsection{Constraints from experiments}
We impose two important constraints from $B\to X_s \gamma$ and
$B_s\to \mu^+\mu^-$. Considering the theoretical uncertainties, we
take $2.0\times 10^{-4} < {\rm Br}(B\to X_s \gamma)< 4.5\times
10^{-4}$, as generally analysed in literatures.
Phenomenologically, Br($B\to X_s \gamma$) directly constrains
$|C_{7\gamma}(m_b)|^2 + |C^\prime_{7\gamma}(m_b)|^2$ at the
leading order. Due to the strong enhancement factor
$m_{\tilde{g}}/m_b$ associated with single $\delta^{LR(RL)}_{23}$
insertion term in $C^{(\prime)}_{7\gamma}(m_b)$,
$\delta^{LR(RL)}_{23}$ ($\sim 10^{-2}$) are more severely
constrained than $\delta^{LL(RR)}_{23}$. However, if the
left-right mixing of scalar bottom quark $\delta^{LR}_{33}$ is
large ($\sim 0.5$), $\delta^{LL(RR)}_{23}$ is constrained to be
order of $10^{-2}$ since the double insertion term
$\delta^{LL(RR)}_{23} \delta^{LR(LR*)}_{33}$ is also enhanced by
$m_{\tilde{g}}/m_b$. The branching ratio $B_s \rightarrow \mu^+
\mu^-$ in MSSM is given as
\begin{eqnarray}\label{bsmu}
{\rm Br}(B_s \rightarrow \mu^+ \mu^-) &=& \frac{G_F^2
\alpha^2_{\rm em}}{64 \pi^3} m^3_{B_s} \tau_{B_s} f^2_{B_s}
|\lambda_t|^2 \sqrt{1 - 4 \widehat{m}^2}
[(1 - 4\widehat{m}^2) |C_{Q_1}(m_b) - C^\prime_{Q_1}(m_b)|^2 + \nonumber\\
&& |C_{Q_2}(m_b) - C^\prime_{Q_2}(m_b) + 2\widehat{m}(C_{10}(m_b)
- C^\prime_{10}(m_b) )|^2]
\end{eqnarray}
where $\widehat{m} = m_\mu/m_{B_s}$. In the middle and large
$\tan\beta$ case the term proportional to $(C_{10}-C_{10}^\prime)$
in Eq. (\ref{bsmu}) can be neglected. The current experimental
upper bound of ${\rm Br}(B_s\to \mu^+\mu^-)$ is $2.6\times
10^{-6}$~\cite{bsmu}. To translate into the constraint on
$C_{Q_{11,13}}$, we have
\begin{eqnarray}
\sqrt{|C_{Q_{11}}(m_W)-C_{Q_{11}}^\prime(m_W)|^2 +
|C_{Q_{13}}(m_W)-C_{Q_{13}}^\prime(m_W)|^2}\lsim 0.1
\end{eqnarray} Because the bound constrains
$|C_{Q_i}-C_{Q_i}^\prime|$ (i=1, 2),
\footnote{$C_{Q_{1,2}}^{(\prime)}$ are the Wilson coefficients of
the operators $Q_{1,2}^{(\prime)}$ which are Higgs penguin induced
in leptonic and semileptonic B decays and their definition can be
found in Ref.~\cite{hy}. By substituting the quark-Higgs vertex
for the lepton-Higgs vertex it is straightforward to obtain Wilson
coefficients relevant to hadronic B decays.} we can have values of
$|C_{Q_i}|$ and $|C_{Q_i}^\prime|$ larger than those in
constrained MSSM (CMSSM) with universal boundary conditions at the
high scale and scenarios of the extended minimal flavor violation
in MSSM~\cite{kane} in which $|C_{Q_i}^\prime|$ is much smaller
than $|C_{Q_i}|$. At the same time we require that predicted Br of
$B\to X_s \mu^+\mu^-$ falls within 1 $\sigma$ experimental bounds.

We also impose the current experimental lower bound $\Delta M_s >
14.4 ps^{-1}$~\cite{msd} and experimental upper bound ${\rm Br}
(B\to X_s g)< 9\%$~\cite{bsg}. Because $\delta^{LR(RL)}_{23}$ is
constrained to be order of $10^{-2}$ by Br($B \to X_s \gamma$),
their contribution to $\Delta M_s$ is small. The dominant
contribution to $\Delta M_s$ comes from $\delta^{LL(RR)}_{23}$
insertion with both constructive and destructive effects compared
with the SM contribution, where the too large destructive effect
is ruled out, because SM prediction is only slightly above the
present experiment lower bound.

As pointed out in section II, due to the gluino-sbottom loop
diagram contribution and the mixing of NHB induced operators onto
the chromomagnetic dipole operator, the Wilson coefficients
$C_{8g}^{(\prime)}$ can be large, which might lead to a too large
Br of $B\to X_s g$. So we need to impose the constraint from
experimental upper bound ${\rm Br} (B\to X_s g)< 9\%$. A numerical
analysis for $C_{8g}^\prime$=0 has been performed in
Ref.\cite{hk}. We carry out a similar analysis by setting both
$C_{8g}$ and $C_{8g}^\prime$ non-zero.

\subsection{ Numerical results for $B\rightarrow \phi K_S$}

In numerical analysis we fix $m_{\tilde g}= m_{\tilde q}= 400 {\rm
GeV}$ and $\tan\beta=30$. We vary the NHB masses in the ranges of
$91 {\rm GeV} \leq m_h \leq 135 {\rm GeV}, 91 {\rm GeV} \leq m_H
\leq 200 {\rm GeV}$ with $m_h < m_H$ and $200 {\rm GeV} \leq m_A
\leq 250 {\rm GeV}$ for the fixed mixing angle $\alpha=0.6, \pi/2$
of the CP even NHBs and scan $\delta^{dAB}_{23}$ in the range
$|\delta^{dAB}_{23}| \leq 0.05$ for A=B and 0.01 for $A\neq B$ (A
= L, R). 

\begin{figure}
{\includegraphics[width=4cm] {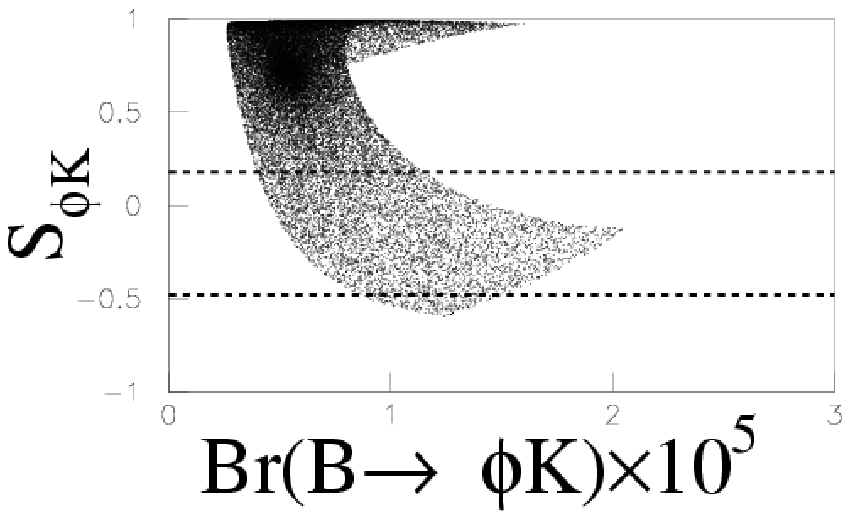}}
{\includegraphics[width=4cm] {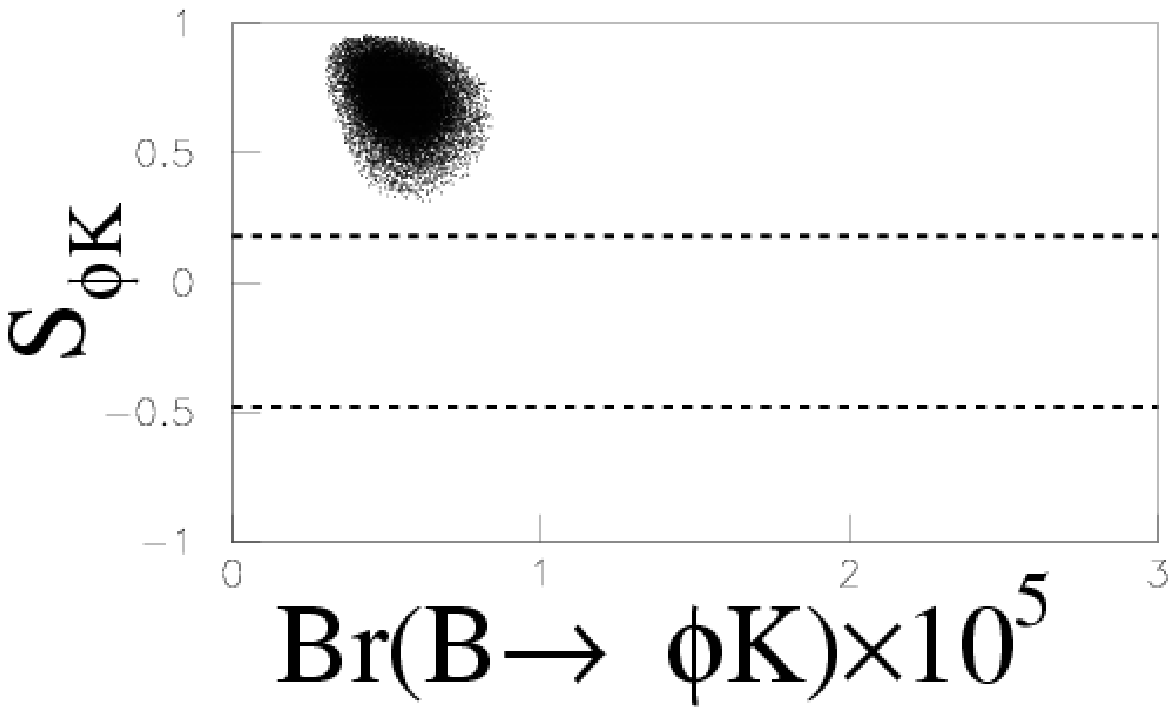}}
{\includegraphics[width=4cm] {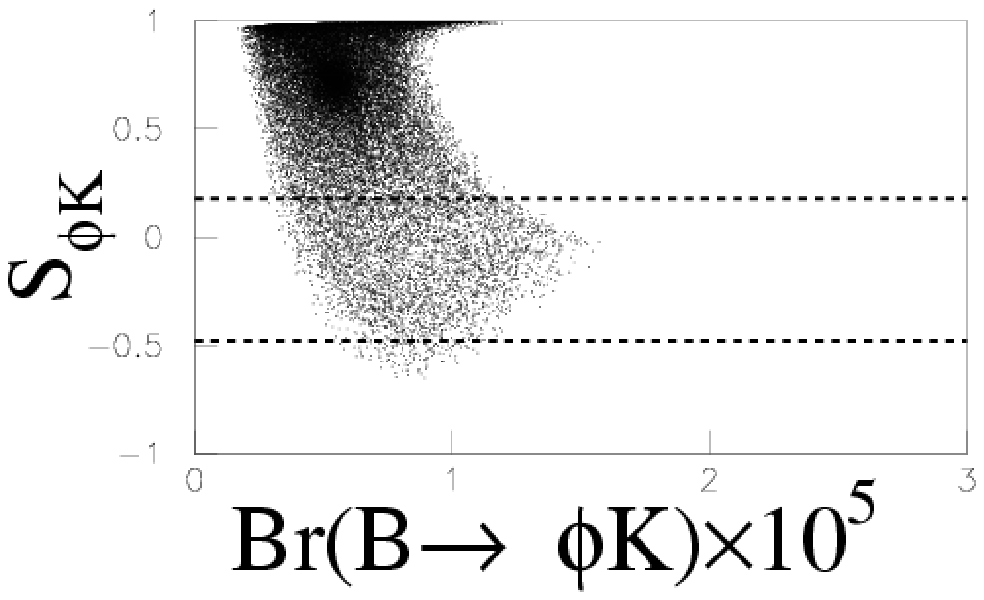}}
\caption{ \label{fig1}
  The correlation between $S_{\phi K_S}$ and Br($B\to
\phi K_S$) for the insertion of only one kind of chirality. (a) is
for the LR insertion, (b) is for the LL insertion with only SM and
NHB contributions included, and (c) is for the LL insertion with
the all contributions included. Current $1\sigma$ bounds are shown
by the dashed lines. }
\end{figure}

\begin{figure}
{\includegraphics[width=4cm] {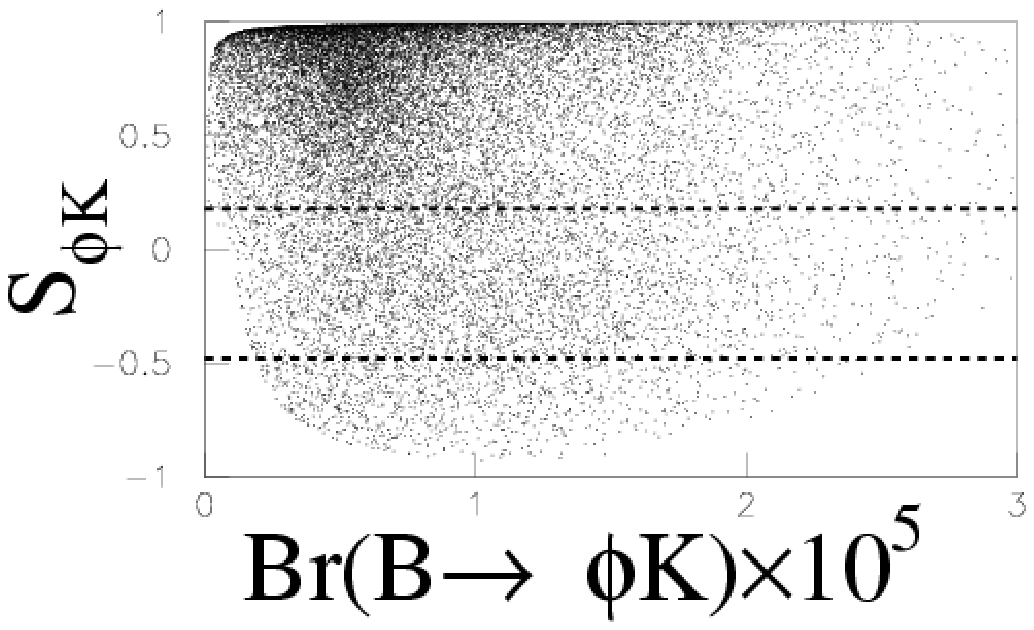}}
{\includegraphics[width=4cm] {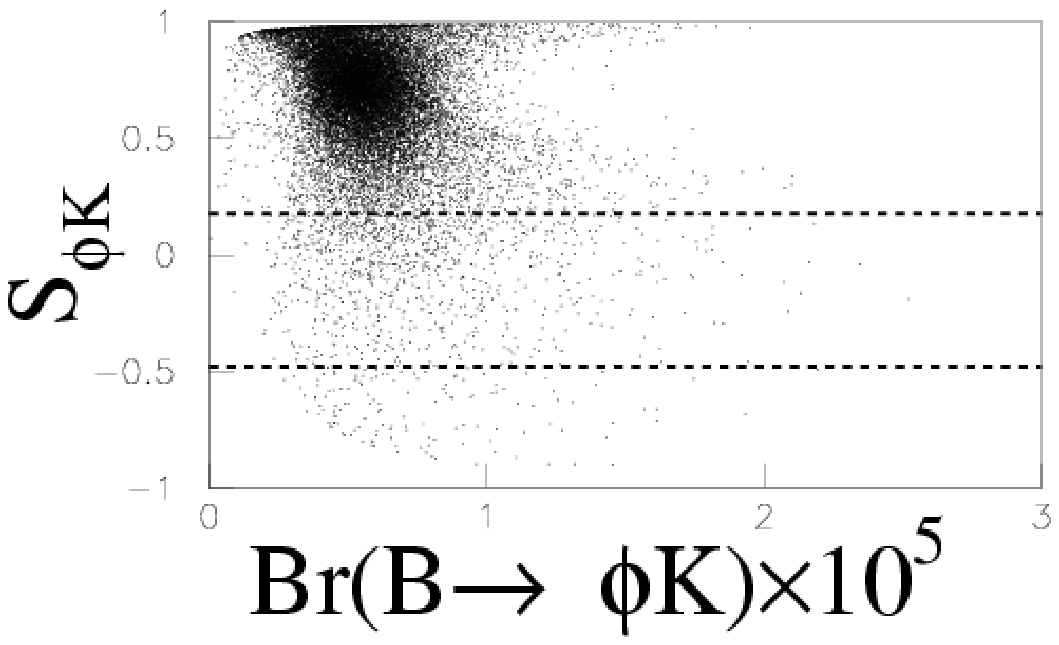}}
{\includegraphics[width=4cm] {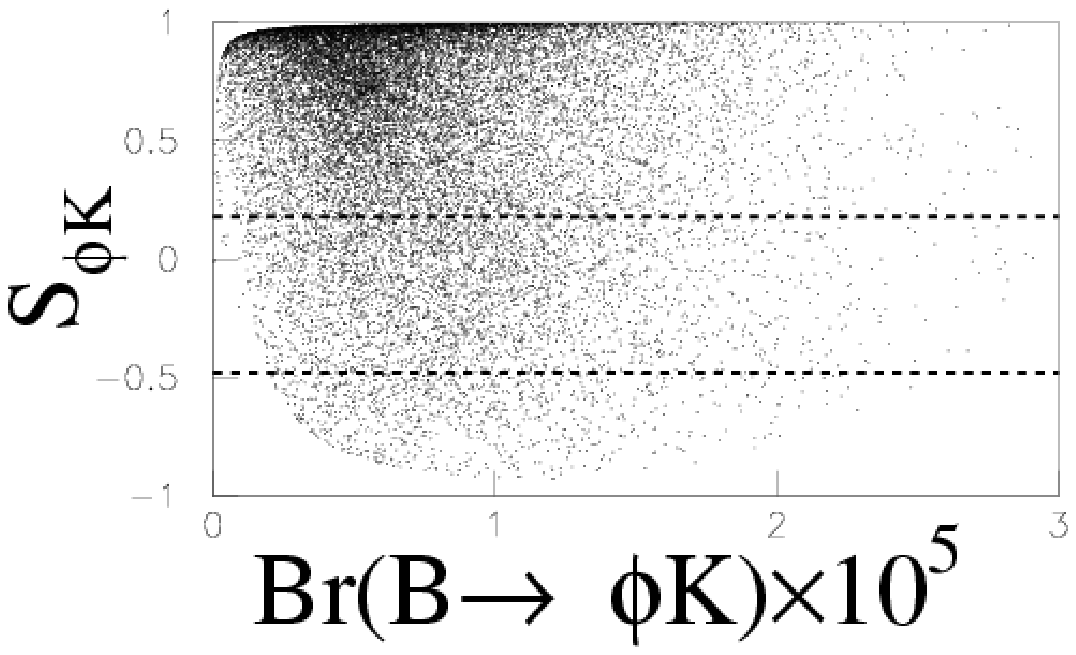}}
\caption{ \label{fig2}
  The correlation between $S_{\phi K_S}$ and Br($B\to
\phi K_S$). (a) is for the LR and RL insertions, (b) is for the LL
and RR insertions with only SM and NHB contributions included, and
(c) is for the LL and RR insertions with the all contributions
included. Current $1\sigma$ bounds are shown by the dashed lines.
}\end{figure}
 Numerical results for $B\rightarrow \phi K_S$ are shown in
{\bf Figs}. \ref{fig1}, \ref{fig2} where the correlation between
$S_{\phi K_S}$ and Br($B\to \phi K_S$) is plotted. {\bf Fig}.
\ref{fig1} is for the insertion of only one kind of chirality.
\ref{fig1}(a), \ref{fig1}(b) and \ref{fig1}(c) correspond to the
LR insertion, the LL insertion with only NHB and SM contributions
included, and the LL insertion with the all contributions
included, respectively. We find that for the LR insertion ({\bf
Fig}. \ref{fig1}(a)) and the LL insertion ({\bf Fig}.
\ref{fig1}(c)) there are regions of parameters where $S_{\phi
K_S}$ falls in $1 \sigma$ experimental bounds and Br is smaller
than $1.6\times 10^{-5}$. In the case of the LL insertion with
only NHB and SM contributions included ({\bf Fig}. \ref{fig1}(b))
$S_{\phi K_S} \geq 0.3$ because $C_i^{\prime}=0$, i=11,...,16 and
consequently $C_i$, i=11,...,16 can not be as large as in the case
with both the LL and RR insertions due to the constraint from
$B_s\to \mu^+\mu^-$. The similar result is obtained for only a RL
insertion or a RR insertion. In {\bf Fig}. \ref{fig2},
\ref{fig2}(a), \ref{fig2}(b), and \ref{fig2}(c) correspond to the
LR and RL insertions, the LL and RR insertions with only NHB and
SM contributions included, and the LL and RR insertions with the
all contributions included, respectively. We find that there are
regions of parameters where $S_{\phi K_S}$ falls in $1 \sigma$
experimental bounds and Br is smaller than $1.6\times 10^{-5}$ in
all three cases. The difference among the three cases is that the
size of the regions is different: the smallest ones are for the LL
and RR insertions with only NHB and SM contributions included, and
the biggest ones for the LL and RR insertions with the all
contributions included, as expected. In the cases of the LL and RR
insertions and the LR and RL insertions $S_{\phi K_S}$ can reach
$-0.9$ or so. Comparing \ref{fig1}(a) \ref{fig1}(c) with
\ref{fig2}(a) \ref{fig2}(c), one can see that the case with both
the LR and RL insertions (both the LL and RR insertions) has
parameter regions with negative $S_{\phi K_S}$ larger than those
in the case with the LR (LL) insertion.

\subsection{ Numerical results for $B\rightarrow \eta^\prime K_S$}

\begin{figure}
{\includegraphics[width=4cm] {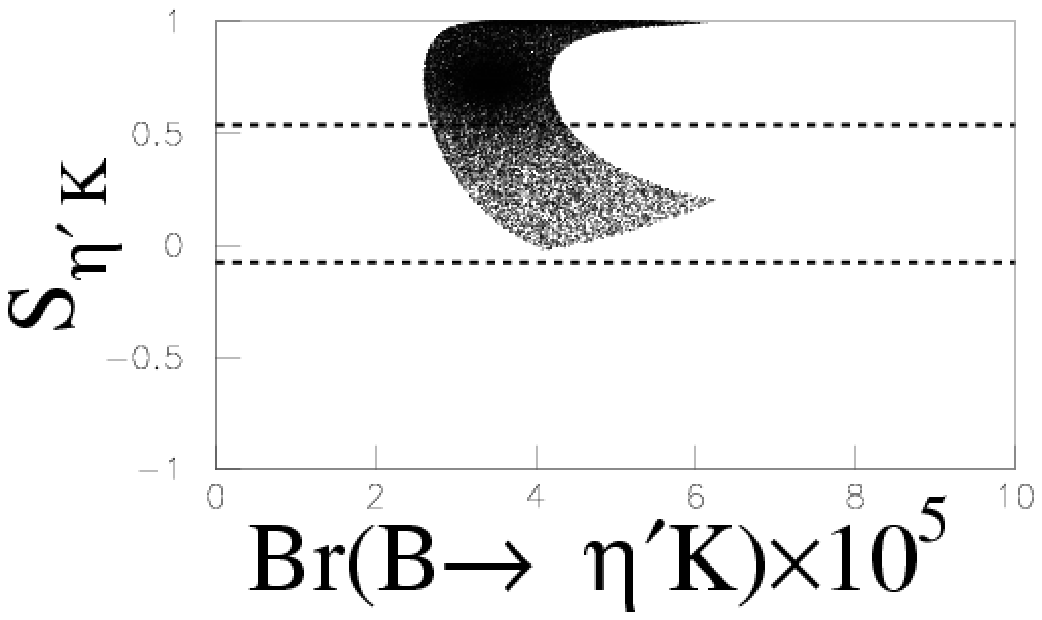}}
{\includegraphics[width=4cm] {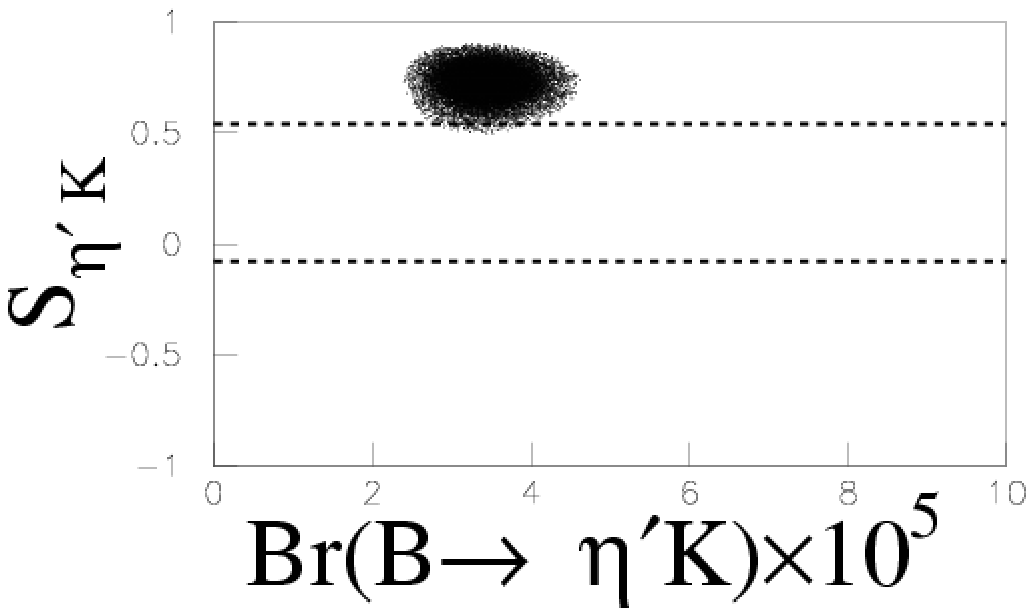}}
{\includegraphics[width=4cm] {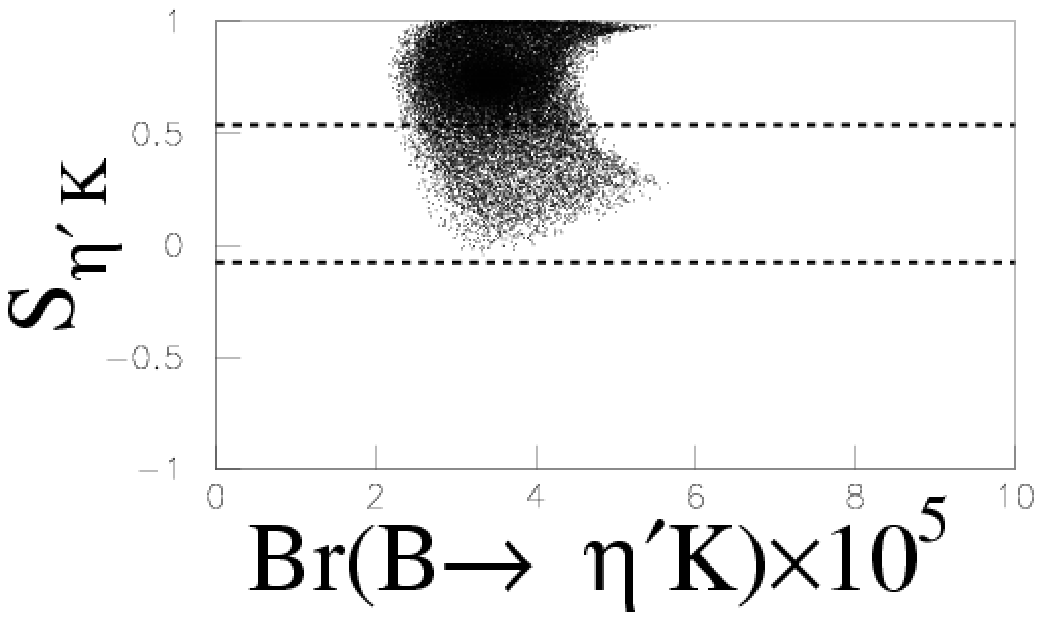}}
\caption{ \label{fig3}
  The correlation between $S_{\ep K_S}$ and Br($B\to
\ep K_S$) for the insertion of only one kind of chirality. (a) is
for the LR insertion, (b) is for the LL insertion with only SM and
NHB contributions included, and (c) is for the LL insertion with
the all contributions included. Current $1\sigma$ bounds are shown
by the dashed lines. }
\end{figure}

\begin{figure}
{\includegraphics[width=4cm] {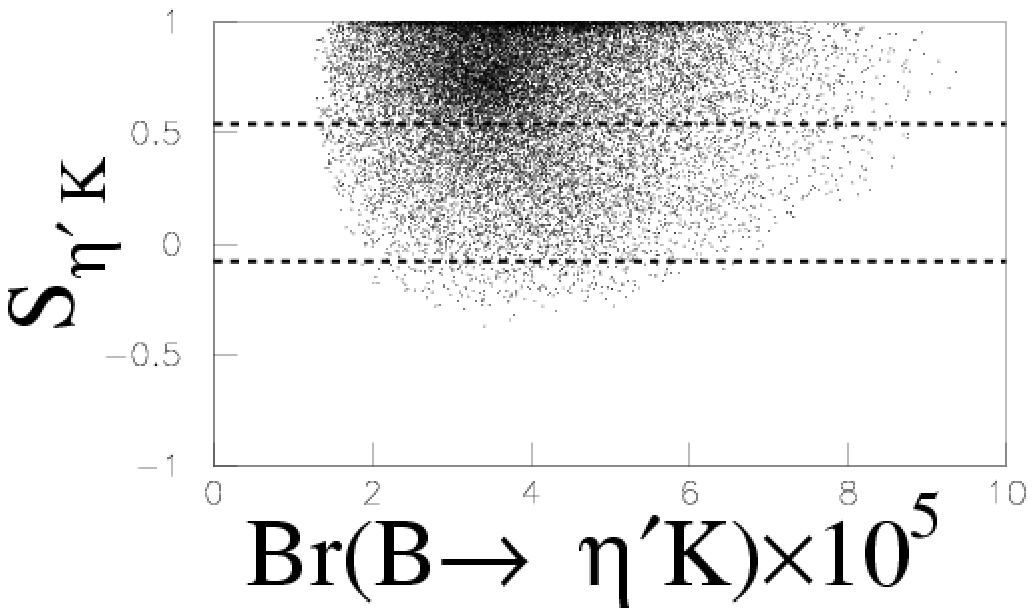}}
{\includegraphics[width=4cm] {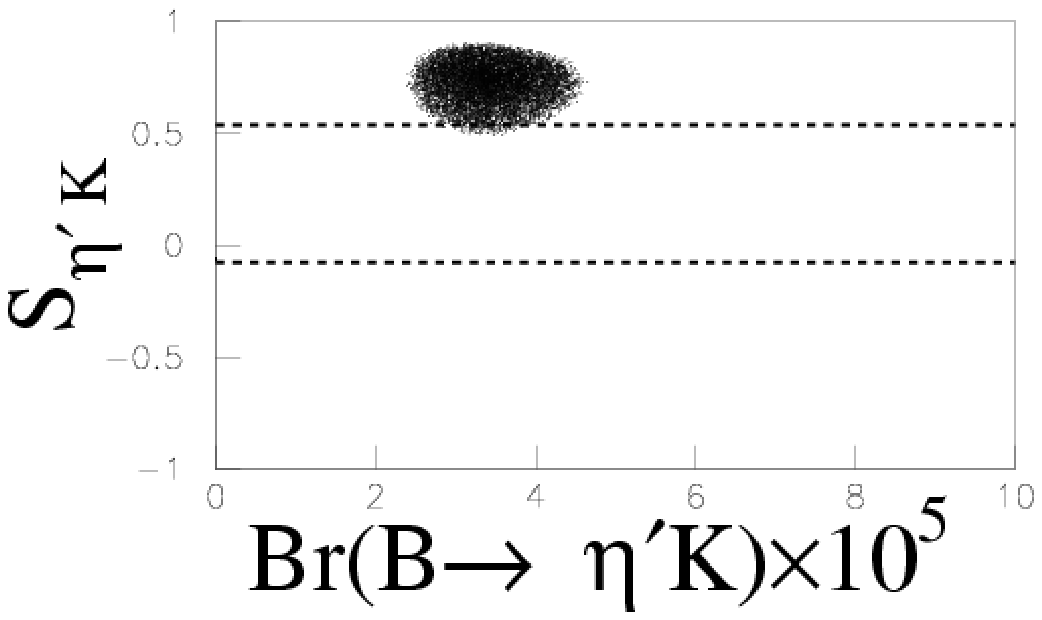}}
{\includegraphics[width=4cm] {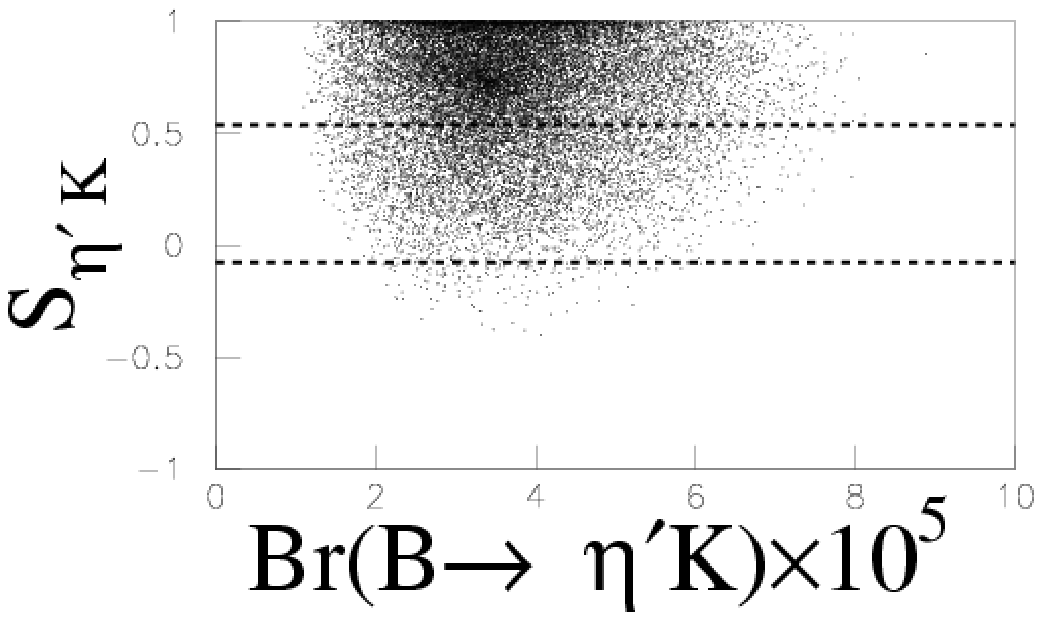}}
\caption{ \label{fig4}
  The correlation between $S_{\ep K_S}$ and Br($B\to
\ep K_S$). (a) is for the LR and RL insertions, (b) is for the LL
and RR insertions with only SM and NHB contributions included, and
(c) is for the LL and RR insertions with the all contributions
included. Current $1\sigma$ bounds are shown by the dashed lines.}
\end{figure}

Input parameters used in this subsection are the same as those in
last subsection. Numerical results for $B\rightarrow \ep K_S$ are
shown in {\bf Figs}. \ref{fig3}, \ref{fig4} where the correlation
between $S_{\ep K_S}$ and Br($B\to \ep K_S$) is plotted. The {\bf
Fig}. \ref{fig3} is for the insertion of only one kind of
chirality. \ref{fig3}(a), \ref{fig3}(b), and \ref{fig3}(c)
correspond to the LR insertion, the LL insertion with only NHB and
SM contributions included, and the LL insertion with the all
contributions included, respectively. In all three cases $S_{\ep
K_S}\geq 0$ and in the case of the LL insertion with only NHB and
SM contributions included the minimal value of $S_{\ep K_S}$ is
0.5. The similar result is obtained for only a RL insertion or a
RR insertion. In {\bf Fig}. \ref{fig4}, \ref{fig4}(a),
\ref{fig4}(b), and \ref{fig4}(c) correspond to the LR and RL
insertions, the LL and RR insertions with only NHB and SM
contributions included, and the LL and RR insertions with the all
contributions included, respectively. We find that there are
regions of parameters where $S_{\ep K_S}$ falls in $1 \sigma$
experimental bounds and Br is larger than $2.7\times 10^{-5}$ and
smaller than $11\times 10^{-5}$ in the first and third cases. For
the first case, i.e., the case of the LR and RL insertions, there
is most of region of parameters where $S_{\ep K_S}$ is positive.
In the third case, i.e., the case of the LL and RR insertions with
the all contributions included, the region which corresponds to
positive $S_{\ep K_S}$ increases. For the second case, i.e., the
case of the LL and RR insertions with only NHB and SM
contributions included, $S_{\ep K}\geq 0.5$ in all regions of
parameters, in contrast to the cases of the LR and RL insertions
and the LL and RR insertions with the all contributions included.
The reason is that the new contributions come mainly from the
combination $C_{13}(m_W)-C_{13}^\prime(m_W)$ which is constrained
by $B_s\to \mu^+\mu^-$. It is the same reason that make {\bf Fig}.
\ref{fig4}(b) similar to {\bf Fig}. \ref{fig3}(b). We would like
to note that our results on Br($B\to \ep K_S$) can agree with the
experimental measurement \cite{pdg} \be Br(B\to \eta^\prime
K)=(5.8^{+ 1.4}_{-1.3}) \times 10^{-5} \ee Indeed, it has been
shown that the data can be explained within theoretical
uncertainties in SM without introducing any new
mechanism\cite{bn}.

\begin{figure}
{\includegraphics[width=4cm] {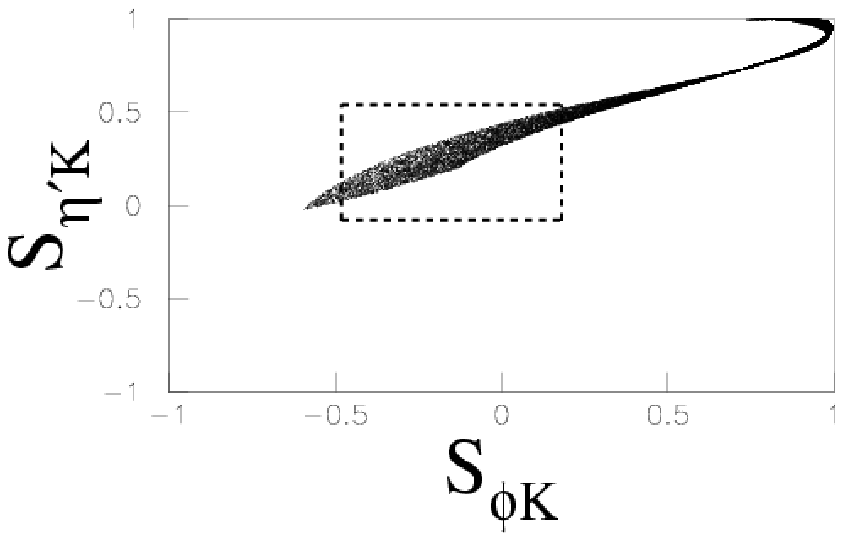}}
{\includegraphics[width=4cm] {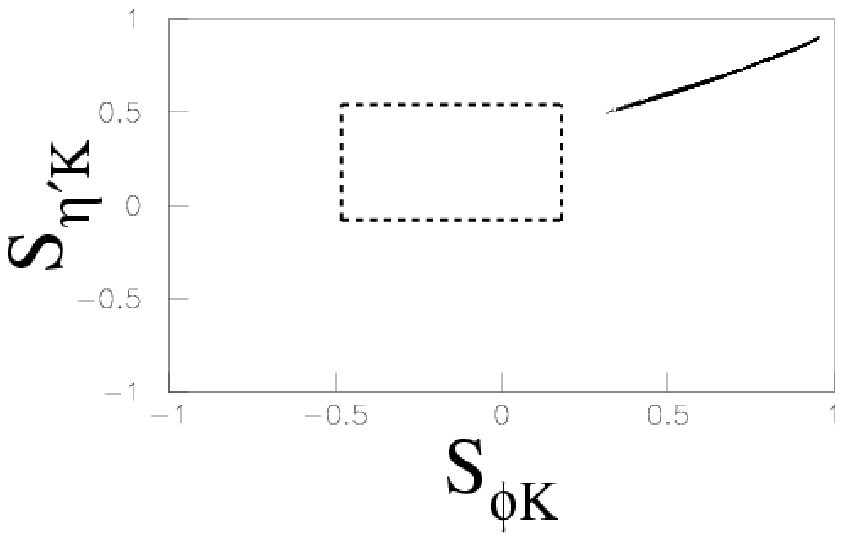}}
{\includegraphics[width=4cm] {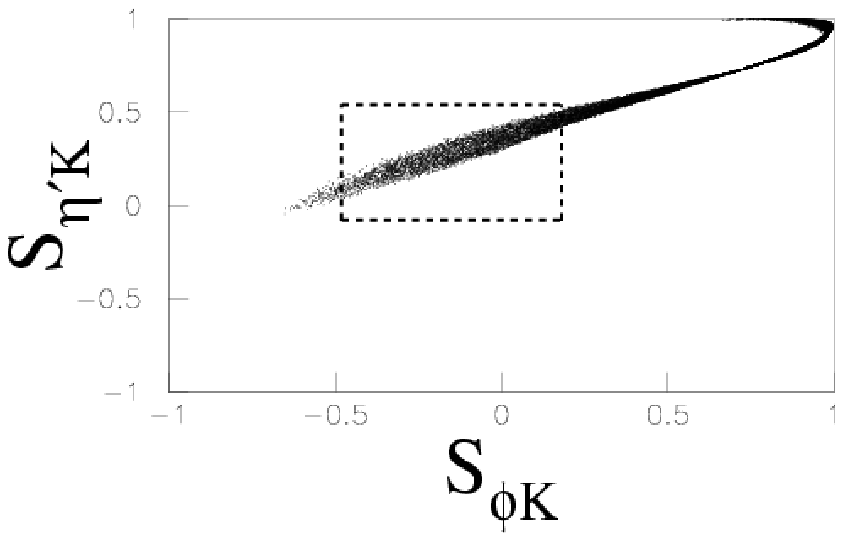}}
\caption{ \label{fig5}
  The correlation between $S_{\ep K_S}$ and $S_{\phi K_S}$ for the
insertion of only one kind of chirality. (a) is for the LR
insertion, (b) is for the LL insertion with only SM and NHB
contributions included, and (c) is for the LL insertion with the
all contributions included. Current $1\sigma$ bounds are shown by
the dashed lines. }
\end{figure}

\begin{figure}
{\includegraphics[width=4cm] {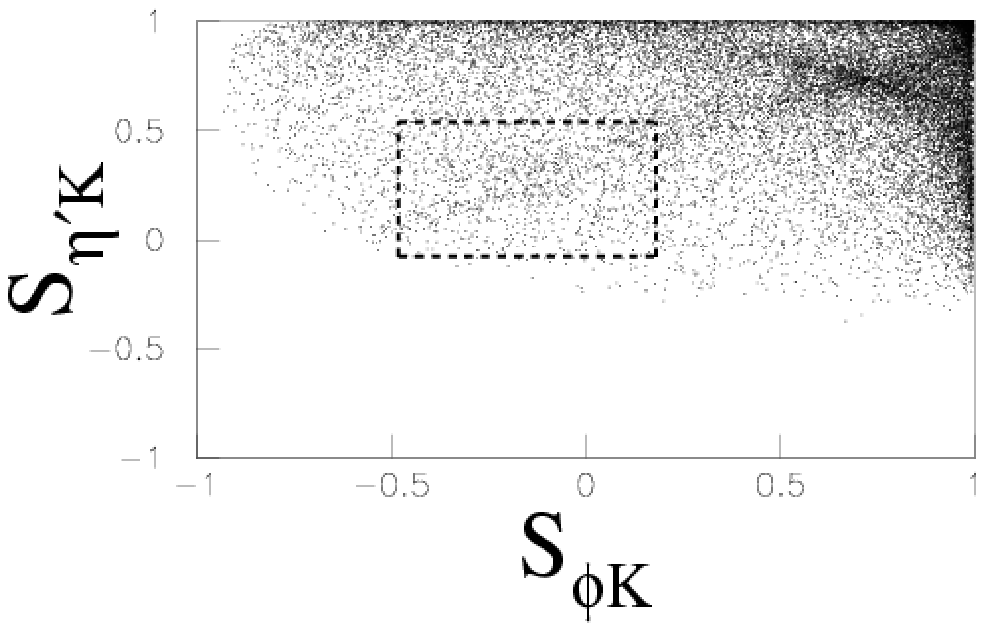}}
{\includegraphics[width=4cm] {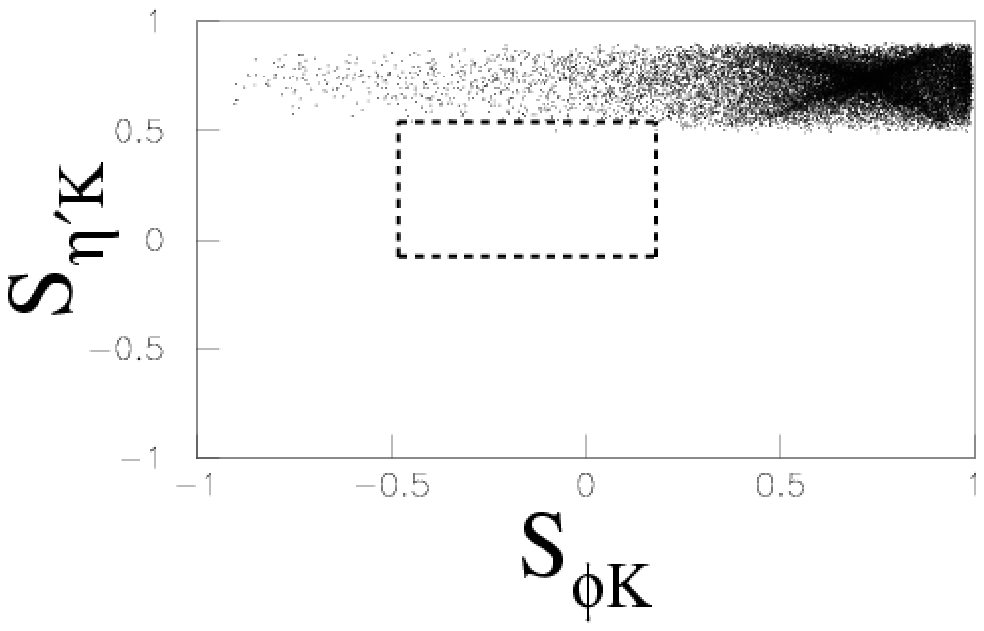}}
{\includegraphics[width=4cm] {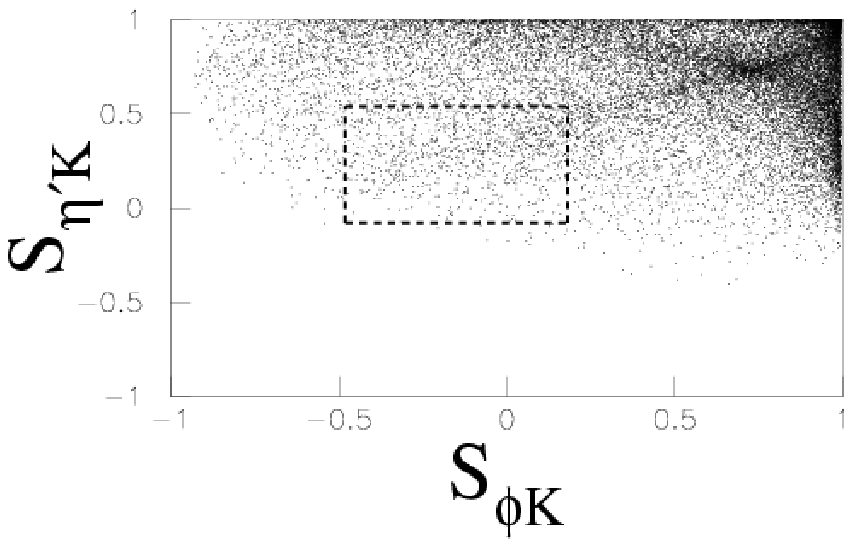}}
\caption{ \label{fig6}
  The correlation between $S_{\ep K_S}$ and $S_{\phi K_S}$. (a) is
for the LR and RL insertions, (b) is for the LL and RR insertions
with only SM and NHB contributions included, and (c) is for the LL
and RR insertions with the all contributions included. Current
$1\sigma$ bounds are shown by the dashed lines.}
\end{figure}

In order to show explicitly there is the region of parameters in
which $S_{\ep K_S}$ is positive and $S_{\phi K_S}$ is negative for
a set of values of parameters we plot the correlation between
$S_{\phi K_S}$ and $S_{\ep K_S}$ in {\bf Figs}. \ref{fig5},
\ref{fig6}. The {\bf Fig}. \ref{fig5} is devoted to the case with
an insertion of only one kind of chirality. {\bf Fig}.
\ref{fig5}(a) is for the LR insertion, {\bf Fig}. \ref{fig5}(b)
and {\bf Fig}. \ref{fig5}(c) are for only NHB and SM contributions
included and the all contributions included, respectively, with
the LL insertion. One can see from {\bf Fig}. \ref{fig5}(a) that
there is the region in which $S_{\ep K_S}$ is positive and
$S_{\phi K_S}$ is negative and their minimal values are 0 and
$-0.6$, respectively. {\bf Fig}. \ref{fig5}(b) shows that with
only the LL insertion there is no region in which $S_{\ep K_S}$ is
positive and $S_{\phi K_S}$ is negative if one includes only NHB
and SM contributions. The reason is that $C_{11,13}^\prime(m_W)=0$
without a RR insertion (see, eg. (\ref{wilson})) so that the
Br($B_s\to \mu^+\mu^-$ ) upper bound limits the size of
$C_{11,13}$, which leads to $S_{\phi K_S}\geq 0.3$, as shown in
{\bf Fig}. \ref{fig1}(b). When switching on all SUSY contributions
among which the contribution coming from the chromo-magnetic
dipole operator is dominant because its Wilson coefficient can be
large due to including the double insertions (see, Eq.
(\ref{wilson}) ), there is the region in which $S_{\ep K_S}$ is
positive and $S_{\phi K_S}$ is negative and their minimal values
are 0 and $-0.7$, respectively, as shown in {\bf Fig}.
\ref{fig5}(c). We would like to point out that our result in the
case with only an LL insertion is different from that in
Ref.\cite{kkou} because we include the $\alpha_s$ corrections of
hadronic matrix elements and the effects of NHB induced operators.
We have also calculated the correlation between $S_{\phi K_S}$ and
$S_{\ep K_S}$ in the case with only an RL insertion and an RR
insertion respectively. In the case with only an RL insertion,
although there is the region in which $S_{\ep K_S}$ is positive
and $S_{\phi K_S}$ is negative but there is no region in which
both $S_{\ep K_S}$ and $S_{\phi K_S}$ are in agreement with data
in the $1 \sigma$ bound. In the case with only an RR insertion,
there are some points in the parameter space for which both
$S_{\ep K_S}$ and $S_{\phi K_S}$ are in agreement with data in the
$1 \sigma$ bound.

 {\bf Fig}. \ref{fig6} is for (a) the LR
and RL insertions, (b) only NHB and SM contributions included and
(c) the all contributions included with the LL and RR insertions.
We can see from the figure that the region exists for all three
cases. For the case of the LR and RL insertions, there is a small
region in which both $S_{\ep K_S}$ and $S_{\phi K_S}$ are
negative. For the case of the LL and RR insertions with only NHB
and SM contributions included, the region of parameters in which
$S_{\ep K}$ is positive and $S_{\phi K_S}$ is negative is smaller
than that in the case of the LR and RL insertions and there is no
region in which both $S_{\ep K_S}$ and $S_{\phi K_S}$ are
negative. There is almost no region in which $S_{\ep K_S}$ is
smaller than $-0.5$ for the case of the LL and RR insertions with
the all contributions included and also for the case of LR and RL
insertions.

The numerical results are obtained for
$m_{\tilde{g}}=m_{\tilde{q}}$=400 GeV. For smaller gluino and
squark masses, the Wilson coefficient $C_{8g}^{(\prime)}$ becomes
larger, which could make the effect on $S_{MK}$ and Br larger.
However, indeed the effect is limited due to the constraint from
$B\to X_s g$. For fixed $m_{\tilde{g}}$, the Wilson coefficient
$C_{8g}^{(\prime)}$ is not sensitive to the variation of the mass
of squark in the range about from 100 GeV to 1.5 Tev. Therefore,
the numerical results are not sensitive to the squark mass and
would have a sizable change when the gluino mass decreases. For
the very big gluino and squark masses (say, several TeV), SUSY
effects drop, i.e., one reaches the decoupling limit.

\section{Conclusions and Discussions}

In summary we have calculated the Wilson coefficients at LO for
the new operators which are induced by NHB penguins using the MIA
with double insertions in the MSSM. We have calculated the
$\alpha_s$ order hadronic matrix elements of the new operators for
$B\rightarrow \phi K_S$ and $B\rightarrow \eta^\prime K_S$ in the
QCD factorization approach. Using the Wilson coefficients and
hadronic matrix elements obtained, we have calculated the
time-dependent CP asymmetries $S_{\phi k}$ and $S_{\eta^\prime K}$
and branching ratios for the decays $B\rightarrow \phi K_S$ and
$B\rightarrow \eta^\prime K_S$. It is shown that in the reasonable
region of parameters where the constraints from $B_s-\bar{B}_s$
mixing , $\Gamma(b \to s \gamma)$, $\Gamma(b \to s g)$, $\Gamma(b
\to s\mu^+ \mu^-)$, $B\to \mu^+\mu^-$ are satisfied, the branching
ratio of the decay for $B\rightarrow \phi K_S$ can be smaller than
$1.6 \times 10^{-5}$, and $S_{\phi K_s}$ can be negative. In some
regions of parameters $S_{\phi K_s}$ can be as low as $-0.9$. The
branching ratio and the time dependent CP asymmetry of the decay
for $B\rightarrow \ep K_S$ can agree with experiments within
$1\sigma$ deviation in quite a large region of parameters. In
particular, our result in the case with only an LL insertion or an
RR insertion is different from that in Ref.\cite{kkou} because we
include the $\alpha_s$ corrections of hadronic matrix elements of
operators and the effects of NHB induced operators.

It is necessary to make a theoretical prediction in SM as
precision as we can in order to give a firm ground for finding new
physics. For the purpose, we calculate the twist-3 and weak
annihilation contributions in SM using the method in
Ref.~\cite{ch} by which there is no any phenomenological parameter
introduced. The numerical results show that the annihilation
contributions to Br are negligible, the twist-3 contributions to
Br are also very small, smaller than one percent, and both the
annihilation and twist-3 contributions to the time-dependent CP
asymmetry are negligible. The conclusion remains in MSSM and we
have neglected the annihilation contributions in numerical
calculations.

In conclusion, we have shown that the recent experimental
measurements on the time-dependent CP asymmetry in $B\to \phi K_S$
and $B\to \ep K_S$, which can not be explained in SM, can be
explained in MSSM if there are flavor non-diagonal squark mass
matrix elements of second and third generations whose size
satisfies all relevant constraints from known experiments ($B\to
X_S\gamma, B_s\to \mu^+\mu^-, B\to X_s \mu^+\mu^-, B\to X_s g,
\Delta M_s$, etc.). Therefore, if the present experimental results
remain in the future, it will signal the significant breakdown of
the standard model and that MSSM is a possible candidate of new
physics.
\section*{Acknowledgement}
The work was supported in part by
the National Nature Science Foundation of China.
XHW is supported by KOSEF Sundo Grant R02-2003-000-10085-0 and
the China Postdoctoral Science Foundation. One of the authors(JFC) are
grateful to Professor C.-D. L$\ddot{\rm u}$ and Y.-D. Yang for their
helpful comments and discussions.

\section*{Note added}When completing the paper we received the
reference~\cite{datanew}. In Ref.~\cite{datanew} the new bound of
$Br(B_s\to \mu^+\mu^-)$ is given as
$$
Br(B_s\to \mu^+\mu^-) < 5.8 \times 10^{-7}~ ~~~ {\rm at ~~90\%~~
confidence~~ level}.
$$
The new bound would give a more stringent constraint on the
contributions of NHBs. We have carried out preliminary
calculations. The preliminary results show that it is still
possible that NHB contributions to $S_{MK}$ ($M=\phi,
\eta^\prime$) alone can make a sizable even significant deviation
from the SM. In other words, the qualitative conclusion in the
paper is still valid if using the new bound.

\section*{\bf Appendix A~~ Loop functions}
In this Appendix, we present the one-loop function of Wilson
coefficients in this work.
\begin{eqnarray}
b_{1,2}(x) &=& x \frac{\partial B_{1,2}(x,y)}{\partial x}|_{y \to x} \nonumber\\
p_{1,2}(x) &=& x \frac{\partial C_{1,2}(x,y)}{\partial x}|_{y \to x} \nonumber\\
F_{2,12,4,34}(x) &=& x \frac{\partial f_{2,12,4,34}(x)}{\partial x} \nonumber\\
b^\prime_{1,2}(x) &=& \frac{x^2}{2}
 \frac{\partial^2 B_{1,2}(x,y)}{\partial x^2}|_{y \to x} \nonumber\\
p^\prime_{1,2}(x) &=& \frac{x^2}{2}
 \frac{\partial^2 C_{1,2}(x,y)}{\partial x^2}|_{y \to x} \nonumber\\
F^\prime_{2,12,4,34}(x) &=& \frac{x^2}{2}
 \frac{\partial^2 f_{2,12,4,34}(x)}{\partial x^2} \nonumber\\
f^\prime_b(x) &=& \frac{x^2}{2}
 \frac{\partial^2 f_{b0}(x)}{\partial x^2}
\end{eqnarray}
with
\begin{eqnarray}
f_{12}(x) = 9 f_1(x) + f_2(x), \hspace{1cm} f_{34}(x) = 9 f_3(x) +
f_4(x)/2 \nonumber
\end{eqnarray}
where $B_{1,2}(x,y)$ and $C_{1,2}(x,y)$ are defined in
Ref.~\cite{murayama0212} and $f_{1,2,3,4,b0}$ in Ref.~\cite{hw}.
\section*{\bf Appendix B~~ Coefficients $a_i$}
We shall give the explicit expressions of coefficients $a_i$ of
the matrix element of the effective Hamiltonian in SM
\cite{bbns1,bn}which are not given in the content. For the
integral which contains end-point singularity, we give a corrected
one by including transverse momentum effects of partons and the
Sudakov factor. The coefficients $a_i$ can generally be divided
into two parts $a_i(M_1M_2)=a_{i,\rm I}(M_1M_2)+a_{i,\rm
II}(M_1M_2)$:
\begin{eqnarray}\label{ai}
\begin{array}{l@{\qquad}l@{\,}l}
\vspace{0.2cm}
   \displaystyle{a_{1,\rm I} = C_1 + \frac{C_2}{N_c} \left[ 1
    + \frac{C_F\alpha_s}{4\pi}\,V_{M_2} \right]} ,
   &
    \displaystyle{a_{1,\rm II} = \frac{C_2}{N_c}\,\frac{C_F\pi\alpha_s}{N_c}\,
    H} ,&
    \nonumber\\
\vspace{0.2cm}
   \displaystyle{ a_{2,\rm I} = C_2 + \frac{C_1}{N_c} \left[ 1
    + \frac{C_F\alpha_s}{4\pi}\,V_{M_2} \right]} ,
   &
    \displaystyle{ a_{2,\rm II}
    = \frac{C_1}{N_c}\,\frac{C_F\pi\alpha_s}{N_c}\,
    H} , &
    \nonumber\\
\vspace{0.2cm}
   \displaystyle{ a_{3,\rm I} = C_3 + \frac{C_4}{N_c} \left[ 1
    + \frac{C_F\alpha_s}{4\pi}\,V_{M_2} \right]} ,
   &
    \displaystyle{ a_{3,\rm II} = \frac{C_4}{N_c}\,\frac{C_F\pi\alpha_s}{N_c}\,
    H },&
    \nonumber\\
\vspace{0.2cm}
   \displaystyle{ a_{4,\rm I}^p = C_4 + \frac{C_3}{N_c} \left[ 1
    + \frac{C_F\alpha_s}{4\pi}\,V_{M_2} \right]
    + \frac{C_F\alpha_s}{4\pi}\,\frac{P_{M_2,2}^p}{N_c}} ,
   &
    \displaystyle{ a_{4,\rm II} = \frac{C_3}{N_c}\,\frac{C_F\pi\alpha_s}{N_c}\,
    H }, & \nonumber\\
\vspace{0.2cm}
   \displaystyle{ a_{5,\rm I} = C_5 + \frac{C_6}{N_c} \left[ 1
    + \frac{C_F\alpha_s}{4\pi}\,(-V_{M_2}') \right]} , &
    \displaystyle{ a_{5,\rm II} = \frac{C_6}{N_c}\,\frac{C_F\pi\alpha_s}{N_c}\,
    \left( -H\right) }& , \\
\vspace{0.2cm}
   \displaystyle{ a_{6,\rm I}^p = C_6 + \frac{C_5}{N_c} \left( 1
    - 6\cdot\frac{C_F\alpha_s}{4\pi} \right)
    + \frac{C_F\alpha_s}{4\pi}\,\frac{P_{M_2,3}^p}{N_c}} ,
   &
    \displaystyle{ a_{6,\rm II} = 0 ,}& \\
\vspace{0.2cm}
   \displaystyle{ a_{7,\rm I} = C_7 + \frac{C_8}{N_c} \left[ 1
    + \frac{C_F\alpha_s}{4\pi}\,(-V_{M_2}') \right]} ,
   &
    \displaystyle{ a_{7,\rm II} = \frac{C_8}{N_c}\,\frac{C_F\pi\alpha_s}{N_c}\,
    (-H)}
    , &\\
\vspace{0.2cm}
   \displaystyle{ a_{8,\rm I}^p = C_8 + \frac{C_7}{N_c} \left( 1
    - 6\cdot\frac{C_F\alpha_s}{4\pi} \right)
    + \frac{\alpha}{9\pi}\,\frac{P_{M_2,3}^{p,{\rm EW}}}{N_c}} ,
   &
    \displaystyle{ a_{8,\rm II} = 0} ,&\\
\vspace{0.2cm}
   \displaystyle{ a_{9,\rm I} = C_9 + \frac{C_{10}}{N_c} \left[ 1
    + \frac{C_F\alpha_s}{4\pi}\,V_{M_2} \right]} ,
   &
    \displaystyle{ a_{9,\rm II}
             = \frac{C_{10}}{N_c}\,\frac{C_F\pi\alpha_s}{N_c}\,
    H} , &\\
\vspace{0.2cm}
   \displaystyle{ a_{10,\rm I}^p = C_{10} + \frac{C_9}{N_c} \left[ 1
    + \frac{C_F\alpha_s}{4\pi}\,V_{M_2} \right]
    + \frac{\alpha}{9\pi}\,\frac{P_{M_2,2}^{p,{\rm EW}}}{N_c}} ,
   &
    \displaystyle{ a_{10,\rm II}
       = \frac{C_9}{N_c}\,\frac{C_F\pi\alpha_s}{N_c}\,
    H} \mbox{。} &
\end{array}
\end{eqnarray}
where $M_2$ means the "emission" meson, which is $\phi$ in the
process $B\to K\phi$ and is $\eta'$ or $K$ in the process $B\to
\eta' K$. The vertex contributions in Eq. (\ref{ai}) are given as
following:
\begin{eqnarray}\label{FM}
   V_{M_2} &=& 12\ln\frac{m_b}{\mu} - 18 + \int_0^1\! dx\,g(x)\,\Phi_{M_2}(x)
    \,, \nonumber\\
   V_{M_2}' &=& 12\ln\frac{m_b}{\mu} - 6 + \int_0^1\! dx\,g(1-x)\,\Phi_{M_2}(x)
    \,, \nonumber\\
   g(x) &=& 3\left( \frac{1-2x}{1-x}\ln x-i\pi \right) \nonumber\\
   &&\mbox{}+ \left[ 2 \,\mbox{Li}_2(x) - \ln^2\!x + \frac{2\ln x}{1-x}
    - (3+2i\pi)\ln x - (x\leftrightarrow 1-x) \right] ,
\end{eqnarray}
 The constants $18$ and $6$
are specific to the NDR scheme. Next, the penguin contributions are
\begin{eqnarray}\label{PK}
   P_{M_2,2}^p &=& C_1 \left[ \frac43\ln\frac{m_b}{\mu}
    + \frac23 - G_{M_2}(s_p) \right]
    + C_3 \left[ \frac83\ln\frac{m_b}{\mu} + \frac43
    - G_{M_2}(0) - G_{M_2}(1) \right] \nonumber\\
   &&\mbox{}+ (C_4+C_6) \left[ \frac{4n_f}{3}\ln\frac{m_b}{\mu}
    - (n_f-2) G_{M_2}(0) - G_{M_2}(s_c) - G_{M_2}(1) \right] \nonumber\\
   &&\mbox{}- 2 C_{8g}^{\rm eff} \int_0^1 \frac{dx}{1-x}\,
    \Phi_{M_2}(x) \,, \nonumber\\
   P_{M_2,2}^{p,{\rm EW}} &=& (C_1+N_c C_2) \left[
    \frac43\ln\frac{m_b}{\mu} + \frac23 - G_{M_2}(s_p) \right]
    - 3\,C_{7\gamma}^{\rm eff} \int_0^1 \frac{dx}{1-x}\,\Phi_{M_2}(x) \,,
\end{eqnarray}
where $n_f=5$ is the number of light quark flavours, and $s_u=0$,
$s_c=(m_c/m_b)^2$ are mass ratios involved in the evaluation of the
penguin diagrams. The small contributions from  electroweak penguin operators
are neglected in $P^p_{M_2,2}$ within our approximations.
 Similar comments apply to (\ref{hatPK}) below. The function $G_{M_2}(s)$ is given by
\begin{eqnarray}\label{GK}
   G_{M_2}(s) &=& \int_0^1\!dx\,G(s-i\epsilon,1-x)\,\Phi_{M_2}(x) \,, \\
   G(s,x) &=& -4\int_0^1\!du\,u(1-u) \ln[s-u(1-u)x] \,.
\end{eqnarray}
Compared with the twist-2 terms, the twist-3 terms do not have factor
$1-x$, which is cancelled in the calculation. We therefore find
\begin{eqnarray}\label{hatPK}
   P_{M_2,3}^p &=& C_1 \left[ \frac43\ln\frac{m_b}{\mu}
    + \frac23 - \hat G_{M_2}(s_p) \right]
    + C_3 \left[ \frac83\ln\frac{m_b}{\mu} + \frac43
    - \hat G_{M_2}(0) - \hat G_{M_2}(1) \right] \nonumber\\
   &+& (C_4+C_6) \left[ \frac{4n_f}{3}\ln\frac{m_b}{\mu}
    - (n_f-2) \hat G_{M_2}(0) - \hat G_{M_2}(s_c) - \hat G_{M_2}(1) \right]
    - 2 C_{8g}^{\rm eff} \,, \nonumber\\
   P_{M_2,3}^{p,{\rm EW}} &=& (C_1+N_c C_2) \left[
    \frac43\ln\frac{m_b}{\mu} + \frac23 - \hat G_{M_2}(s_p) \right]
    - 3\,C_{7\gamma}^{\rm eff} \,,
\end{eqnarray}
with
\begin{equation}\label{penfunction1}
   \hat G_{M_2}(s) = \int_0^1\!dx\,G(s-i\epsilon,1-x)\,\Phi_p^{M_2}(x) \,.
\end{equation}

The hard-scattering contributions $H_{M_1M_2}$ are defined as following:
\begin{eqnarray}
H_{M_1\phi}\propto \int d\xi du dv \left[ {\phi_B \left (\xi\right)\over \xi}
                  {\phi_{M_1} \left(u\right)\over u}
          {\phi_{M_2}\left(v\right)\over v}+
 {2 \mu_{M_1}\over m_B}{\phi_B\left(\xi\right)\over \xi}
                     {{\phi_\sigma\left(u\right)\over 6}\over u^2}
             {\phi_{M_2}\left(v\right)\over v} \right]. \label{HH}
\end{eqnarray}
We assume
\begin{eqnarray}\label{phikp}
&&\Phi_B(x) = N_B x^2(1-x)^2 \exp \left[ -{m_B^2 x^2\over 2
\omega_B^2}\right]
\end{eqnarray} with normalization factor $N_B$ satisfying $\int^1_0 dx \Phi_B(x)
= 1$, which is the popular used form in literature. Fitting
various B decay data, $\omega_B$ is determined to be 0.4 GeV
\cite{li}. With the Sudakov effects to cancel end-point
singularity, Eq.(\ref{HH}) turns into
\begin{eqnarray}
H_{M_1\phi} &\propto&
\int d\xi du dv\, d^2{\bf k_\perp} d^2 {\bf k_{1\perp}} d^2 {\bf k_{2\perp}}
\nonumber\\
 &\  \times& \Bigg[\ \ {{-um_B^4}
 \phi_B(\xi)\phi_{M_1}(u)\phi_{M_2}(v)\over {[\xi
 um_B^2+({\bf k_\perp }-{\bf k_{1\perp}})^2][-uvm_B^2+({\bf k_\perp}-{\bf
 k_{1\perp}}+{\bf k_{2\perp}})^2]} }
\nonumber\\
&&\ +\
{ 2\mu_{{M_1}} m_B^5 u v
 \phi_B(\xi)\frac{\phi_{\sigma}(u)}{6}\phi_{M_2}(v)\over [\xi
 um_B^2+({\bf k_\perp }-{\bf k_{1\perp}}^2)][-uvm_B^2+({\bf k_\perp}-{\bf
 k_{1\perp}}+{\bf k_{2\perp}})^2]^2} \Bigg] \,.
\end{eqnarray}
Calculating  the upper integral in $b$ space,  we obtain
the hard spectator scattering contribution
\begin{eqnarray}\label{hbpi}
H_{M_1\phi}&=&\frac{4\pi^2}{N_c}\frac{f_{M_1} f_B}{F^{B\rightarrow{M_1}}_+(m^2_\phi) m^2_B}
\int dxdydz
\int bdb b_2db_2 {\mathscr P}_B(z,b){\mathscr P}_{M_2}(x,b_2)\nonumber\\
&&\mbox{} \times
\bigg\{- u m_B^4 {\mathscr P}_{{M_1}}(y,b)
K_0(v b_2)\times \left[\theta(b_2-b)I_0(ub)
 K_0(u b_2)  + \theta(b-b_2)I_0(ub_2)K_0(u b)
\right]\nonumber\\
&&\hspace{0.9cm} +2uv\mu_{{M_1}}m_B^5
\frac{{\mathscr P}_{\sigma}(y,b)}{6}
 \frac{b_2}{2v}K_{-1}(vb_2)\times\left[\theta(b_2-b)I_0(ub)K_0(u b_2)+
 \theta(b-b_2)I_0(u b_2)K_0(u b)  \right] ~\bigg\}\nonumber\\
\
\end{eqnarray}
where $K_i$, $I_i$~are modified Bessel functions of order $i$~and ${\mathscr
P}_B$, ${\mathscr P}_{M_1}$, ${\mathscr P}_{M_2}$~are $B$, $M_1$, $M_2$~corrected meson
amplitudes with the exponentials $S_B$, $S_{M_2}$~and
$S_{M_1}$~respectively~\cite{li, li1}。

\section*{\bf Appendix C.~~ Implementation of $\eta$--$\eta'$
mixing} \label{sec:mix}

Though in the calculation of weak decay amplitudes we only
consider the case with an $\eta^{\prime}$ meson in the final
state, we have to include the effect from $\eta$--$\eta'$ mixing.
In the following discussions we use the FKS scheme to describe
the mixing between $\eta$--$\eta'$\cite{fks, bn}. The matrix
elements  of local operators evaluated between the vacuum and
$\eta^{(\prime)}$,
 of the flavor-diagonal axial-vector and pseudoscalar
current densities, read
\begin{equation}\label{deffh}
\begin{array}{l@{}l@{\,}l}
\vspace{0.1cm}
   \langle P(q)|\bar q\gamma^\mu\gamma_5 q|0\rangle
   &= -\displaystyle{\frac{i}{\sqrt2}}\,f_P^q\,q^\mu \,, \qquad &
   2 m_q \langle P(q)|\bar q\gamma_5 q|0\rangle
   = -\displaystyle{\frac{i}{\sqrt2}}\,h_P^q \,, \\
   \langle P(q)|\bar s\gamma^\mu\gamma_5 s|0\rangle
   &= -i f_P^s\,q^\mu \,, &
   2 m_s \langle P(q)|\bar s \gamma_5 s|0\rangle
   = -i h_P^s \,,
\end{array}
\end{equation}
where $q=u$ or $d$. We assume exact isospin symmetry and identify
$m_q\equiv\frac12(m_u+m_d)$. We also need the anomaly matrix
elements
\begin{equation}\label{anomalyme}
   \langle P(q)|\frac{\alpha_s}{4\pi}\,G_{\mu\nu}^A\,
   \widetilde{G}^{A,\mu\nu}|0\rangle = a_P \,,
\end{equation}
where we use the convention
\begin{equation}
   \widetilde{G}^{A,\mu\nu} = -\frac12\,\epsilon^{\mu\nu\alpha\beta}
   G_{\alpha\beta}^A \qquad (\epsilon^{0123}=-1)
\end{equation}
for the dual field-strength tensor. In all cases $P=\eta$ or
$\eta'$ denotes the physical pseudoscalar meson state.

The Eq.(\ref{deffh}) and Eq.(\ref{anomalyme}) including
 ten non-perturbative parameters $f_P^i$,
$h_P^i$, and $a_P$, which however are not all independent, connect
by
\begin{equation}
   \partial_\mu(\bar q\gamma^\mu\gamma_5 q) = 2im_q\,\bar q\gamma_5 q
   - \frac{\alpha_s}{4\pi}\,G_{\mu\nu}^A\,\widetilde{G}^{A,\mu\nu}
\end{equation}
(and similarly with $q$ replaced by $s$) yields four relations
between the various parameters, which can be summarised as
\begin{equation}\label{div}
   a_P = \frac{h_P^q-f_P^q\,m_P^2}{\sqrt2} = h_P^s-f_P^s\,m_P^2 \,.
\end{equation}
which leaves us with six independent parameters now.

The physical states are related to the flavor states in the FKS
scheme by The
\begin{equation}
   \left( \begin{array}{c}
    |\eta\rangle \\ |\eta'\rangle
   \end{array} \right)
   = \left( \begin{array}{cc}
    \cos\phi & ~-\sin\phi \\
    \sin\phi & \phantom{~-}\cos\phi
   \end{array} \right)
   \left( \begin{array}{c}
    |\eta_q\rangle \\ |\eta_s\rangle
   \end{array} \right) ,
\end{equation}
where $|\eta_q\rangle=(|u\bar u\rangle+|d\bar d\rangle)/\sqrt2$
and $|\eta_s\rangle=|s\bar s\rangle$, then the same mixing angle
applies to the decay constants $f_P^i$ and $h_P^i$ with the
normalization given by (\ref{deffh}). We can re-define the decay
constants in terms of $f_{s,q}$ and a mixing angle $\phi$ as
\begin{equation}
\begin{array}{l@{}l@{\,}l}
\vspace{0.1cm}
   f_\eta^q &= f_q\cos\phi \,, \qquad &
   f_\eta^s = -f_s\sin\phi \,, \\
   f_{\eta'}^q &= f_q\sin\phi \,, &
   f_{\eta'}^s = f_s\cos\phi \,,
\end{array}
\end{equation}
and an analogous set of equations for the $h_P^i$. Inserting these
results into (\ref{div}) allows us to express all ten
non-perturbative parameters in terms of the decay constants $f_q$,
$f_s$ and the mixing angle $\phi$. We obtain
\begin{eqnarray}\label{hqres}
   h_q &=& f_q\,(m_\eta^2\,\cos^2\phi + m_{\eta'}^2\,\sin^2\phi)
    - \sqrt2 f_s\,(m_{\eta'}^2-m_\eta^2)\,\sin\phi\cos\phi \,, \\
   h_s &= &f_s\,(m_{\eta'}^2\,\cos^2\phi + m_\eta^2\,\sin^2\phi)
    - \frac{f_q}{\sqrt2}\,(m_{\eta'}^2-m_\eta^2)\,\sin\phi\cos\phi \,,
\end{eqnarray}
and
\begin{eqnarray}\label{theas}
   a_\eta &=& -\frac{1}{\sqrt2}\,(f_q\,m_\eta^2-h_q)\,\cos\phi
    = -\frac{m_{\eta'}^2-m_\eta^2}{\sqrt2}\,\sin\phi\cos\phi\,
    (-f_q\,\sin\phi+\sqrt2 f_s\,\cos\phi) \,, \\
  a_{\eta'} &= &-\frac{1}{\sqrt2}\,(f_q\,m_{\eta'}^2-h_q)\,\sin\phi
   = -\frac{m_{\eta'}^2-m_\eta^2}{\sqrt2}\,\sin\phi\cos\phi\,
   (f_q\,\cos\phi+\sqrt2 f_s\,\sin\phi) \,.
\end{eqnarray}

\end{document}